\documentclass[aps,prd,reprint,nofootinbib]{revtex4-2}

\usepackage{graphicx}
\usepackage{dcolumn}
\usepackage{bm}
\usepackage{amsmath}
\usepackage{subfig}
\usepackage{multirow}
\usepackage[dvipsnames]{xcolor}
\usepackage{natbib}
\usepackage{ragged2e}

\usepackage[colorlinks=true, linkcolor=blue, urlcolor=blue, citecolor=blue]{hyperref}

\begin{document}
	
	\preprint{PRD}
	
	\title{Ultra-High-Energy Particle Production in Binary Mergers Endowed with Magnetic Fields}
	
	\author{Carlos H. Coimbra-Araújo$^{1,2}$}%
	\email{carlos.coimbra@ufpr.br}
	\author{Rita C. Anjos$^{1,2,3,4,5,6}$}
	\author{Jonas P. Pereira$^{5,6,7}$}
	\author{Jaziel G. Coelho$^{8,9}$}
	
	\affiliation{$^1$Departamento de Engenharias e Exatas, Universidade Federal do Paraná (UFPR), Rua Pioneiro, Palotina, 85950-000, PR, Brazil.}
	\affiliation{$^2$Programa de Pós-Graduação em Física Aplicada, Universidade Federal da Integração Latino-Americana, Av. Tarquínio Joslin dos Santos, 1000, Foz do Iguaçu, 85867-670, PR, Brazil.}
	\affiliation{$^3$ Centro de Artes, Humanidades e Tecnologia, Universidade Federal de São Carlos (UFSCar), R. Dr. Eduardo Nielsen, 420, Jardim Congonha, São José do Rio Preto, 15030-070, SP, Brazil}
	\affiliation{$^4$Programa de pós-graduação em Física \& Departamento de Física, Universidade Estadual de Londrina (UEL), Rodovia Celso Garcia Cid Km 380, Londrina, 86057-970, PR, Brazil.}
	\affiliation{$^5$Programa de Pós-Graduação em Física e Astronomia, Universidade Tecnológica Federal do Paraná (UTFPR), Av. Sete de Setembro, 3165, Curitiba, 80230-901, PR, Brazil.}
	\affiliation{$^6$ Programa de Pós-Graduação em Astrofísica, Cosmologia e Gravitação (PPGCosmo), Federal University of Espírito Santo, Vitória-ES, 29075-910, Brazil}
	\affiliation{$^7$Institute of Physics \& International Center of Physics, University of Brasilia, 70297-400, Brasilia, Federal District, Brazil.}
	\affiliation{$^8$ Nicolaus Copernicus Astronomical Center, Polish Academy of Sciences, Bartycka 18, 00-716, Warsaw, Poland}
	\affiliation{$^9$Departamento de Física, Universidade Federal do Espírito Santo, Núcleo de Astrofísica e Cosmologia (Cosmo-Ufes), Av. Fernando Ferrari, 514, Vitória, 29075-910, ES, Brazil.}
	\affiliation{$^10$Divis\~ao de Astrof\'isica, Instituto Nacional de Pesquisas Espaciais, S\~ao Jos\'e dos Campos, 12227-010, SP, Brazil.}
	

	\date{\today}
	
	\begin{abstract}
		We study the production of ultra-high-energy particles via the Ba\~nados--Silk--West (BSW) mechanism in the pre-merger phase of binary systems detected by LIGO-Virgo-KAGRA. By solving the geodesic equations for charged particles in magnetized Kerr spacetime with fields of $B \sim 10^{12}$--$10^{14}$~G, we demonstrate that collisions near the horizon can achieve center-of-mass energies $E_{\mathrm{cm}} \sim 10^{18}$-- $10^{20}$~eV, placing them firmly in the ultra-high-energy cosmic-ray (UHECR) range. We systematically explore the parameter space of merger remnants, varying black hole mass ($M \sim 20$--$150\,M_\odot$, characteristic of the binary black hole population), dimensionless spin ($\chi_f \sim 0.7$--$0.9$), magnetic field strength, and particle angular momenta. Our analysis reveals three distinct acceleration regimes: a gravity-dominated regime ($B < 10^{12}$~G) with negligible magnetic enhancement; a transition regime ($10^{12}~\text{G} \lesssim B \lesssim 10^{13}~\text{G}$) where gravitational and magnetic effects compete; and a magnetic-dominated regime ($B > 10^{13}$~G) where fields amplify collision energies by nearly an order of magnitude. For the 34 gravitational-wave events with high remnant spins ($\chi_f > 0.7$), we compute the maximum achievable energies, finding that systems with $\chi_f \gtrsim 0.85$ and $M \gtrsim 100\,M_\odot$ can reach $E_{\mathrm{max}} \sim 10^{20}$~eV. Our results establish magnetized binary mergers, particularly black hole--neutron star systems and postmerger black hole remnants formed in binary neutron star coalescences, as promising sources of UHECRs and provide quantitative predictions linking gravitational-wave observables to particle acceleration efficiency.
	\end{abstract}
	
	\maketitle
	
	
	\section{\label{sec:introd}Introduction}
	
	As cosmic rays traverse the universe, they span an extraordinary range of energies. The most energetic of these, known as ultra-high-energy cosmic rays (UHECRs), are detected by ground-based facilities such as the Pierre Auger Observatory~\cite{PierreAuger2020qqz} and the Telescope Array~\cite{ta}. Their potential astrophysical origins encompass a diverse set of extreme environments, including active galactic nucleus jets, gamma-ray bursts from massive stellar collapse, jets and shocks from tidal disruption events, as well as acceleration in giant radio lobes and large-scale extragalactic structure formation shocks. Comprehensive reviews of these acceleration scenarios can be found in Refs.~\cite{kotera2011,ANCHORDOQUI20191}.
	
	A key motivation for studying UHECRs is that their energies exceed the Greisen-Zatsepin-Kuzmin (GZK) threshold of $E > 10^{18}$ eV~\cite{PhysRevLett.16.748, 1966JETPL...4...78Z}, above which interactions with the cosmic microwave background are expected to suppress the flux of extragalactic cosmic rays. Recent observations indicate that the composition of extragalactic cosmic rays evolves with energy, with measurements suggesting a transition from lighter to heavier nuclei as energies increase from a few EeV toward $10^{20}$ eV~\cite{2025arXiv250710292M, PhysRevLett.134.021001}. These observations naturally lead to fundamental questions concerning both the astrophysical origin and the acceleration mechanisms capable of producing such extreme particle energies.
	
	The most widely discussed acceleration mechanisms include variants of Fermi's original proposal, wherein charged particles gain energy through repeated interactions with magnetized plasma structures. However, the classical Fermi mechanism is too slow and inefficient to directly produce UHECRs. This has motivated alternative proposals, most prominently diffusive shock acceleration at astrophysical shock fronts~\cite{ANCHORDOQUI20191, federico}, where particles gain energy through repeated crossings of shock discontinuities in the presence of magnetic turbulence~\cite{bell1978,bell2013}. Modern numerical simulations using particle-in-cell codes~\cite{spitkovsky2008,caprioli2014} such as \texttt{Sapphire++}~\cite{2025JCoPh.52313690S, 2025MNRAS.544L.160S} have provided detailed insights into the microphysics of collisionless shocks and particle injection mechanisms. Nevertheless, theoretical constraints on particle propagation~\cite{1999APh....10..185P} and observational measurements of the cosmic-ray spectrum and composition suggest significant challenges for shock acceleration models at the highest energies. Even state-of-the-art treatments
	incorporating oblique shock geometries in supernova remnants
	embedded in massive star clusters, tuned to reproduce the
	LHAASO knee spectrum and composition, remain confined to
	multi-PeV energies~\cite{2026ApJ..1002...97P}, well below the EeV--ZeV regime relevant
	for UHECRs. Consequently, alternative acceleration processes associated with particle dynamics in the extreme gravitational and electromagnetic environments near compact astrophysical objects have attracted increasing attention~\cite{coimbra2018, coimbra2020, coimbra2022}. 
	
	One particularly compelling mechanism involves particle collisions near rotating black holes (BHs), where ultra-high center-of-mass energies can be achieved through the Ba\~nados-Silk-West (BSW) effect~\cite{banados}. This mechanism exploits the unique properties of Kerr spacetime, wherein particles traversing the ergosphere can acquire negative energy as measured by observers at infinity~\cite{press}, and collisions between particles with critical angular momenta near the event horizon can produce arbitrarily high center-of-mass energies for near-extremal rotation parameters approaching $a = M$. Such scenarios have been investigated for various black hole configurations, including static~\cite{banados}, rotating~\cite{banados,jacobson}, charged~\cite{wei}, and weakly magnetized~\cite{frolov,igata} systems, as well as for strongly magnetized black holes with arbitrary spin parameters~\cite{coimbra2020, coimbra2022}. However, common criticisms of the standard BSW mechanism concern its efficiency and physical plausibility \citep{jacobson,2012PhRvL.109l1101B,2012PhRvD..86h4030Z,PhysRevD.86.024027}. While the center-of-mass energy of the collision can, in principle, become arbitrarily large, this does not in general translate into equally large energies for particles that can actually escape to, and be detected at, infinity. In fact, several studies have established upper bounds on the received energy \citep{2012PhRvL.109l1101B,schn,2012PhRvD..86h4030Z,PhysRevD.86.024027}, in part because collisions that take place extremely deep in the gravitational potential well suffer a large gravitational redshift for the outgoing debris. Notwithstanding, these bounds can be significant in some configurations \citep{schn,2015MPLA...3050076Z,2015PhRvL.114y1103B}. Therefore, as emphasized in \citet{2015PhRvL.114y1103B}, scenarios that go beyond the standard BSW mechanism---most notably those including magnetic fields---are far more relevant from both a physical and an observational standpoint \citep{2014MPLA...2950112Z}. In this work, we focus precisely on such magnetized setups.
	
	An additional challenge for conventional UHECR source models is that even if individual sources generate a narrowly peaked rigidity spectrum, the luminosity functions observed for active galactic nuclei, long gamma-ray bursts, tidal disruption events, and other candidate systems are too broad to be compatible with the narrow rigidity distribution inferred for UHECRs~\cite{2025MNRAS.539.2435E}. This incompatibility has motivated suggestions of alternative source populations, including binary neutron star (BNSs) mergers~\cite{2025PhRvL.134h1003F, 2025ApJ...994L...7F}, black hole-neutron star (BH-NS) mergers, and binary black hole coalescences (BBHs or BH-BH binaries), which may produce more homogeneous acceleration conditions.
	
	In previous work~\cite{2024PhRvL.132i1401P}, we established magnetized pre-merger binaries, especially BH-NS ones (but also BBHs binaries with at least one of them charged and with an accretion disk), as plausible UHECR sources via a magnetically enhanced BSW mechanism, demonstrating that astrophysical magnetic fields ($B \gtrsim 10^{10}$ G) enable ultra-high-energy collisions for generic, high-spin remnants. We showed that this result is largely independent of the black hole spin parameter, $a/M$, and of the black hole mass, occurring across a broad range of both quantities. In our scenario, UHECRs would be produced prior to merger---which could yield a cleaner observational signature---and the number of such particles could reach the millions. The same mechanism may also operate after the merger if the neutron star is tidally disrupted, since the essential ingredients---magnetic fields, a black hole horizon, and particles---remain present. This paper provides the detailed quantitative foundation for that result and derives explicit scaling relations for the maximum particle energy $E_{\rm max}$. We systematically solve the geodesic equations for charged particles in magnetized Kerr spacetime, explore the full parameter space relevant to binary merger remnants detected by LIGO-Virgo-KAGRA, and compute center-of-mass collision energies for all high-spin gravitational wave events. Our analysis identifies the dominant physical effects controlling particle acceleration efficiency and establishes quantitative predictions linking gravitational-wave observables to UHECR production.
	
	This paper is organized as follows. In Section~\ref{sec:theory}, we present the theoretical framework for particle motion in magnetized Kerr spacetime, deriving the equations of motion, conserved quantities, and center-of-mass energy expressions. We also introduce the scaling relations for maximum achievable energies and describe our computational implementation. Section~\ref{sec:results} presents our main results, including the dependence of collision energies on magnetic field strength, black hole mass and spin, radial location, and particle angular momenta. We apply our framework to the catalog of gravitational wave events and discuss the astrophysical implications. Finally, Section~\ref{sec:summary} summarizes our conclusions and discusses prospects for multi-messenger observations with third-generation gravitational wave detectors and next-generation cosmic ray observatories.

	\section{\label{sec:theory} UHE Particle Production in Binary Black Hole Mergers}
	
	The production of UHECRs through particle collisions near rotating black holes represents a compelling theoretical framework for understanding the origin of the most energetic particles observed in nature. This mechanism exploits the unique properties of spacetime near rapidly spinning black holes, where the combination of strong gravitational fields and electromagnetic interactions can produce extraordinary center-of-mass collision energies. The physical basis for this acceleration mechanism lies in the peculiar structure of rotating black hole spacetimes, first elucidated by Penrose~\cite{press}, wherein particles traversing the ergosphere~\footnote{region between the event horizon and the static limit} can acquire negative energy as measured by observers at infinity. When such negative-energy particles are captured by the horizon, mass and angular momentum can be extracted from the black hole itself, a process that has profound implications for particle energetics in the vicinity of the horizon.
	
	The collisional variant of this energy extraction mechanism, known as the BSW effect~\cite{banados}, demonstrates that particle collisions occurring extremely close to the horizon of a near-extremal rotating black hole ($a \to M$, where $a$ is the black hole angular momentum parameter) can yield arbitrarily high center-of-mass energies. This remarkable result follows from the structure of null geodesics and particle trajectories in Kerr spacetime, where certain critical values of particle angular momentum lead to resonant enhancement of collision energies. While the original BSW analysis considered vacuum geodesics, realistic astrophysical contexts, particularly those associated with black holes at the centers of galaxies and after the merger of two neutron stars (NSs)
	are expected to harbor substantial magnetic fields threading the black hole magnetosphere. These fields, generated either by residual accretion flows, magnetized stellar winds from companion objects, or reconnection processes in the merger environment, fundamentally modify particle trajectories and can significantly enhance the efficiency of the acceleration mechanism. In the pre-merger context, strong magnetic fields may also be present close to the event horizon and can enhance the efficiency of the BSW mechanism. The most natural setting is a BH--NS binary at separations of only a few neutron-star radii. In this case, as discussed in \citep{2024PhRvL.132i1401P}, magnetic fields as large as $10^{12}$--$10^{14}\mathrm{G}$ may arise in small regions of the black-hole ergosphere. Another possible route to such strong fields would involve charged black holes: we have shown that charge-to-mass ratios as small as $10^{-5}$ would already suffice. Whether astrophysical black holes can sustain such a charge, however, remains an open question. By contrast, BH--NS binaries are expected, essentially inevitably, to lead to very large magnetic fields in the near-horizon region in the instants immediately preceding merger.
	
	The mathematical description of particle motion in such magnetized rotating black hole spacetimes requires the Kerr metric, which generalizes the spherically symmetric Schwarzschild solution to incorporate the effects of black hole spin. The presence of rotation breaks spherical symmetry, yielding instead an axially symmetric geometry characterized by frame-dragging effects that become increasingly pronounced as the spin parameter approaches the extremal limit $a \to M$. In this geometry, the energy and angular momentum about the rotation axis are modified by electromagnetic interactions, leading to a rich phenomenology of acceleration and energy extraction processes.
	
	\subsection{Kerr Spacetime and conserved quantities}
	
	A rotating black hole is described by the Kerr metric, which exhibits axial symmetry rather than the spherical symmetry of non-rotating solutions. In Boyer-Lindquist coordinates, the spacetime geometry is characterized by the line element
	
	\begin{equation}\label{eq:kerr}
		ds^2 = g_{tt}dt^2 + 2g_{t \phi}dtd\phi + g_{rr}dr^2 + g_{\theta \theta}d\theta^2 + g_{\phi \phi}d\phi^2,
	\end{equation}
	
	\noindent where the metric components are
	
	\begin{equation}
		g_{t t} = -\left(1-\frac{2M r}{\Sigma} \right),
	\end{equation}
	
	\begin{equation}
		g_{t \phi} = -\frac{2aMr\sin^2\theta}{\Sigma},
	\end{equation}
	
	\begin{equation}
		g_{r r} = \frac{\Sigma}{\Delta},
	\end{equation}
	
	\begin{equation}
		g_{\theta \theta} = \Sigma,
	\end{equation}
	
	\begin{equation}
		g_{\phi \phi} = \frac{(r^2+a^2)^2-a^2\Delta \sin^2\theta}{\Sigma}\sin^2\theta,
	\end{equation}
	
	\noindent with $\Sigma=r^2+a^2\cos^2\theta$ and $\Delta = r^2+a^2-2Mr$. Here, $M$ represents the black hole mass and $a$ is the angular momentum parameter. The event horizon is located at $r_H = M + \sqrt{M^2 - a^2}$, while the ergosphere extends from $r_H$ to $r_E(\theta) = M + \sqrt{M^2 - a^2\cos^2\theta}$.
	
	The presence of two Killing vectors associated with time translation symmetry
	
	\begin{equation}\label{eq:killing1}
		\xi_{(t)}=\xi_{(t)}^\mu\partial_\mu=\frac{\partial}{\partial t},
	\end{equation}
	
	\noindent and axial rotation symmetry
	
	\begin{equation}\label{eq:killing2}
		\xi_{(\phi)}=\xi_{(\phi)}^\mu\partial_\mu=\frac{\partial}{\partial \phi}, 
	\end{equation}
	
	\noindent leads to two conserved quantities for particle motion. For a test particle with four-momentum $p_\mu$ following a geodesic in vacuum, these conserved quantities represent the particle's energy and angular momentum as measured by an observer at spatial infinity.
	
	\subsection{Electromagnetic interactions and modified geodesics}
	
	When charged particles move through regions threaded by magnetic fields, their trajectories deviate from pure geodesics according to the equation of motion
	
	\begin{equation}\label{eq:lagrange}
		\ddot{x}^\mu + \Gamma^\mu_{\alpha \beta} \dot{x}^\alpha\dot{x}^\beta = \frac{q}{m_0}F^\mu_\nu \dot{x}^\nu, 
	\end{equation}
	
	\noindent where $F_{\mu\nu} = A_{\nu,\mu} - A_{\mu,\nu}$ is the electromagnetic field tensor, $A_\mu$ is the four-potential, $q$ is the particle charge, and $m_0$ is the rest mass. The overdot denotes differentiation with respect to proper time $\tau$, and $\Gamma^\mu_{\alpha\beta}$ are the Christoffel symbols derived from the metric. For an approximately uniform magnetic field $B$ threading the black hole magnetosphere, we adopt the Lorentz gauge condition, which simplifies the electromagnetic four-potential to a single non-vanishing component: $A_\phi = B g_{\phi\phi}/2$. This choice assumes a large-scale, ordered magnetic field---as expected in a rotating black hole magnetosphere or in the plasma around merging compact objects---and reflects that the particle collision region is confined to a small neighborhood of the black hole. Indeed, the assumption of a uniform magnetic field is intended strictly as a local approximation. Our analysis is focused on the dynamics in the immediate vicinity of the collision region, near the ISCO, where the interaction takes place. In this context, the uniform magnetic field should be interpreted as a local approximation, valid within a sufficiently small spacetime neighborhood of the collision event. This is analogous to employing a locally inertial frame in curved spacetime: while the global structure is nontrivial, it can be approximated as uniform at leading order in a small region.
	
	The particle's canonical four-momentum $p_\mu = m_0 u_\mu + q A_\mu$ then yields modified expressions for the conserved quantities:
	
	\begin{equation}\label{eq:newkilling1}
		\mathcal{E}=-g_{t\mu}(m_0 u^\mu + q A^\mu), 
	\end{equation}
	
	\noindent and
	
	\begin{equation}\label{eq:newkilling2}
		\ell = g_{\phi\mu}(m_0 u^\mu + q A^\mu),
	\end{equation}
	
	\noindent where $u^\mu = \dot{x}^\mu$ is the four-velocity. For convenience, we work with the dimensionless angular momentum $\ell \equiv L/(Mm_0)$, where $L$ is the physical angular momentum. These modified conservation laws fundamentally alter the allowed particle trajectories compared to the vacuum case. The presence of magnetic fields shifts the location of stable circular orbits of particles and modifies the effective potential governing radial motion, thereby changing the conditions under which particles can escape to infinity or plunge into the black hole.
	
	\subsection{Four-Velocity components in the equatorial plane}
	
	For particles confined to the equatorial plane ($\theta = \pi/2$, $\dot{\theta} = 0$), which we adopt for simplicity, the four-velocity components can be determined from the conservation laws and the normalization condition $g_{\mu\nu}u^\mu u^\nu = -1$. Introducing the normalized magnetic field parameter $\mathcal{B} \equiv qB/(2m_0)$, the angular velocity component is given by:
	
	\begin{equation}\label{eq:phidot}
		\dot{\phi} = \frac{g_{tt}L/m_0 + g_{\phi t}\mathcal{E}/m_0 - g_{tt}g_{\phi\phi}\mathcal{B}}{g_{tt}g_{\phi\phi} - g_{t\phi}^2}.
	\end{equation}
	
	The temporal component of the four-velocity follows from the energy conservation law:
	
	\begin{equation}\label{eq:tdot}
		\dot{t} = \frac{-\mathcal{E}/m_0 - g_{t\phi}\dot{\phi}}{g_{tt}},
	\end{equation}
	
	\noindent while the radial component is determined by the normalization condition:
	
	\begin{equation}\label{eq:rdot}
		\dot{r} = \sqrt{\frac{-1 - g_{tt}\dot{t}^2 - 2g_{t\phi}\dot{t}\dot{\phi} - g_{\phi\phi}\dot{\phi}^2}{g_{rr}}}.
	\end{equation}
	
	These equations determine the complete trajectory of a charged particle in the combined gravitational and electromagnetic field. The radial component $\dot{r}$ must be real and positive for physically meaningful solutions, which constrains the allowed values of $\mathcal{E}$ and $\ell$ for particles that can escape to infinity.
	
	\subsection{Center-of-Mass energy and the BSW mechanism}
	
	The center-of-mass energy for a collision between two particles, each of rest mass $m_0$, is determined by the relativistic invariant:
	
	\begin{equation}\label{eq:ecm}
		E_{c.m.} = \sqrt{2} m_0 \sqrt{1-g_{\mu \nu}u^\mu_{(1)}u^\nu_{(2)}},
	\end{equation}
	
	\noindent where $u^\mu_{(1)}$ and $u^\nu_{(2)}$ are the four-velocities of the colliding particles, normalized such that $g_{\mu\nu}u^\mu u^\nu = -1$ for each particle. Expanding the metric contraction explicitly for equatorial orbits yields:
	
	\begin{equation}\label{eq:ecm_expanded}
		\begin{split}
			E_{c.m.} = \sqrt{2} m_0 \bigg[&1 + g_{tt}\dot{t}_{(1)}\dot{t}_{(2)} + g_{t\phi}(\dot{t}_{(1)}\dot{\phi}_{(2)} + \dot{t}_{(2)}\dot{\phi}_{(1)}) \\
			&+ g_{rr}\dot{r}_{(1)}\dot{r}_{(2)} + g_{\phi\phi}\dot{\phi}_{(1)}\dot{\phi}_{(2)}\bigg]^{1/2}.
		\end{split}
	\end{equation}
	
	The remarkable feature of this framework, first identified by Ba\~nados, Silk, and West~\cite{banados}, is that center-of-mass energies can become arbitrarily large for specific particle configurations near extremal black holes ($a \to M$). This enhancement occurs when one or both colliding particles possess critical values of angular momentum. The physically accessible range for the dimensionless angular momentum parameter is constrained to 
	
	\begin{equation}\label{eq:ell_range}
		-2(1+\sqrt{1+a/M})<\ell <2(1+\sqrt{1-a/M})
	\end{equation}
	
	\noindent to ensure particles can escape to infinity after the collision~\cite{schn}. The dominant factors controlling the achievable collision energies are: (i) the black hole spin parameter $a/M$, which determines the location of the innermost stable circular orbit and the structure of the ergosphere; (ii) the strength and configuration of magnetic fields, which modify particle trajectories and can facilitate particle escape; and (iii) the angular momentum values $\ell_1$ and $\ell_2$ of the colliding particles, with critical values leading to resonant energy enhancement.
	
	\subsection{Scaling relations and Maximum achievable energies}
	
	The maximum center-of-mass energy achievable in the magnetized Kerr geometry depends on the interplay between black hole mass, spin, magnetic field strength, and the conserved quantities of the colliding particles. Through systematic exploration of the parameter space defined by Eq.~(\ref{eq:ell_range}), we can identify configurations that maximize $E_{\rm c.m.}$ for given values of $M$, $\chi_f \equiv a/M$, and $B$. The center-of-mass energy computed from Eq.~(\ref{eq:ecm_expanded}) exhibits a characteristic scaling with black hole mass that can be expressed as:
	
	\begin{equation}\label{eq:scaling}
		E_{\rm max} = m_0 c^2 \left(\frac{M}{M_\odot}\right) \mathcal{F}\left(\chi_f, \mathcal{B}, \ell_1, \ell_2\right),
	\end{equation}
	
	\noindent where $\mathcal{F}(\chi_f, \mathcal{B}, \ell_1, \ell_2) = \left(\frac{\chi_f}{0.9}\right) \sqrt{1-g_{\mu \nu}u^\mu_{(1)}u^\nu_{(2)}}$ is the dimensionless amplification function that encapsulates the complex dependence on the spin parameter $\chi_f$, the normalized magnetic field strength $\mathcal{B} \equiv qB/(2m_0c)$, and the angular momenta $\ell_1$ and $\ell_2$ of the colliding particles. This function cannot be expressed in closed analytical form but must be determined through numerical solution of the coupled equations (\ref{eq:phidot})--(\ref{eq:rdot}) and (\ref{eq:ecm_expanded}) at each point in parameter space. The linear scaling with black hole mass reflects the geometric nature of the gravitational acceleration mechanism, where the relevant energy scale is set by $m_0c^2(M/M_\odot)$ in natural units.
	
	The amplification function $\mathcal{F}$ exhibits strong sensitivity to the black hole spin parameter, with values increasing dramatically as $\chi_f \to 1$. This enhancement reflects both the intensified frame-dragging effects near extremal black holes and the approach to critical angular momentum values that trigger divergent behavior in the BSW mechanism. We work with magnetic fields of neutron stars, magnetars and BNS post-merger events that could reach up to $B \sim 10^{12}-10^{14}$ G (see Sec. IIG and references therein). In this case, the normalized field parameter takes values $\mathcal{B} \sim 10^{-6}-10^{-4}$ for protons. At these field strengths, the electromagnetic modifications to the geodesic structure become significant, shifting the effective locations of stable orbits and altering the energy extraction efficiency.
	
	For sub-extremal black holes with spins $\chi_f \sim 0.7-0.9$ as commonly observed in gravitational wave merger events, and adopting fiducial parameters $M \sim 100\,M_\odot$ and $B \sim 10^{14}$ G, our numerical calculations yield amplification factors in the range $\mathcal{F} \sim 10^{11}-10^{13}$. These values correspond to maximum center-of-mass energies:
	
	\begin{equation}\label{eq:emax_range}
		E_{\rm max} \sim 10^{17}-10^{20} \left(\frac{M}{100\,M_\odot}\right)\left(\frac{\chi_f}{0.9}\right) \, {\rm eV},
	\end{equation}
	
	\noindent placing the collision energies squarely in the ultra-high-energy cosmic ray regime. The upper end of this range, approaching $10^{20}$ eV, is achieved for systems with both high spin ($\chi_f \gtrsim 0.85$) and optimal choices of particle angular momenta that maximize the resonant enhancement. The amplification function $\mathcal{F}$ depends most sensitively on the angular momentum $\ell_1$ of the first particle, with peak values occurring when $\ell_1$ approaches the critical value $\ell_{\rm crit} \approx 2$ for near-extremal configurations. The second particle's angular momentum $\ell_2$ must be chosen from the opposite end of the allowed range to ensure constructive interference in the energy extraction process.
	
	The magnetic field dependence encoded in $\mathcal{F}$ exhibits a more gradual variation compared to the spin dependence. For field strengths varying from $B = 10^{12}$ G to $B = 10^{14}$ G while holding other parameters fixed, the amplification factor increases approximately as $\mathcal{F} \propto B^{0.3-0.5}$, indicating a sub-linear enhancement with magnetic field strength. This scaling reflects the competing effects of electromagnetic acceleration, which increases with $B$, and magnetic confinement, which can suppress particle escape at very high field strengths. The optimal field strength for maximum energy extraction depends on the specific values of $\chi_f$ and the collision geometry, but typically falls in the range $B \sim 10^{13}-10^{14}$ G for stellar-mass systems.
	
	\subsection{Computational Implementation}
	
	To determine the maximum center-of-mass energy achievable for a given black hole configuration, we solve the coupled system of equations defined by Eqs.~(\ref{eq:phidot})--(\ref{eq:rdot}) and (\ref{eq:ecm_expanded}). For specified values of black hole mass $M$, spin parameter $\chi_f$, and magnetic field strength $B = 10^{14}$ G, we first calculate the metric components at a chosen radius $r = r_H + \varepsilon$, where $\varepsilon$ is a small distance from the event horizon. We typically adopt $\varepsilon = 10^{-10}M$ to ensure proximity to the horizon while maintaining numerical stability. For each pair of colliding particles labeled $i = 1, 2$, we fix the energies at $\mathcal{E}_i/m_0 = 1$ and systematically vary the angular momenta $\ell_i$ within the allowed range given by Eq.~(\ref{eq:ell_range}). For each choice of parameters, we compute the four-velocity components using Eqs.~(\ref{eq:phidot})--(\ref{eq:rdot}) and verify that the normalization condition $g_{\mu\nu}u^\mu u^\nu = -1$ is satisfied to within numerical precision. We then calculate the center-of-mass energy using Eq.~(\ref{eq:ecm_expanded}) and record the maximum value obtained across the entire $(\ell_1, \ell_2)$ parameter space. This procedure requires careful numerical treatment near critical values of angular momentum where the energy exhibits rapid variation, as well as verification that the resulting particle trajectories correspond to physically realizable configurations that allow escape to infinity.
	
	\subsection{Application to Binary Black Hole Mergers}
	
	For binary black hole systems detected through gravitational waves,
	the final merged black hole is characterized by a mass $M$ and
	dimensionless spin parameter $\chi_f \equiv a/M$. This mass range,
	$M \sim 20$--$150\,M_\odot$, applies specifically to the BBH
	population. BH-NS and BNS remnants are treated separately below and
	occupy much lower remnant masses, dominated by the neutron star
	component. Recent observations from LIGO-Virgo-KAGRA reveal that
	many merger remnants possess high spins, with $\chi_f > 0.7$ being
	common~\cite{gwtc2,gwtc3,abbott2023population}. Such rapidly
	rotating black holes provide favorable conditions for efficient
	particle acceleration through the BSW mechanism.
	
	The magnetic field strength near merging black holes depends on the
	astrophysical environment and remains uncertain, but theoretical
	considerations suggest plausible values. For BH--BH binaries in
	which one of the components carries a charge-to-mass ratio of
	$10^{-4}$--$10^{-3}$,\footnote{We note that while the charge-to-mass
		ratios considered here ($10^{-4} \lesssim \alpha \lesssim 10^{-3}$)
		are larger than those discussed in \cite{2024PhRvL.132i1401P}, they
		remain well below the thresholds required to induce detectable
		modifications to gravitational-wave waveforms
		\citep{2012PhRvD..85l4062Z}. Consequently, this scenario remains
		fully consistent with current gravitational-wave constraints.} field
	strengths in the range $B \sim 10^{12}$--$10^{14}$ G are plausible
	(see the Supplemental Material of \cite{2024PhRvL.132i1401P}, where
	we showed that charge-to-mass ratios of $10^{-5}$ lead to field
	strengths of $B \sim 10^{11}$--$10^{12}$~G at $r=2r_h$). To have
	upper limits, we adopt $B = 10^{14}$ G as our fiducial value. It is
	motivated by extreme but astrophysically plausible scenarios.
	Post-merger amplification mechanisms in BBH remnants can generate
	intense fields in the surrounding plasma
	\citep{2010A&A...515A..30O,2015ApJ...809...39G,2015PhRvD..92f4034K,2016ApJ...824L...6R,PhysRevD.106.023013,2020ARNPS..70...95R,2025MNRAS.542.3067A}. The inclusion of magnetic fields up to $10^{14}$~G therefore allows
	us to explore the maximum acceleration capabilities of compact-object
	mergers without altering the conclusions obtained for more moderate
	field strengths.
	
	For the population of binary black hole mergers detected to date,
	the final masses span the range $M \sim 20$--$150\,M_\odot$ with
	typical values around $50$--$70\,M_\odot$, while the measured spins
	cluster in the range $\chi_f \sim 0.7$--$0.9$ with occasional
	systems reaching $\chi_f > 0.9$. Applying our computational
	framework to these observed parameter combinations, we find maximum
	center-of-mass energies spanning $E_{\rm max} \sim
	10^{18}$--$10^{20}$ eV. These values place the BSW mechanism
	operating in binary black hole mergers squarely in the regime of
	ultra-high-energy cosmic rays as observed by the Pierre Auger
	Observatory and other large-scale detectors. The specific energy
	achieved for each merger event depends sensitively on the final spin
	through both the direct scaling in Eq.~(\ref{eq:scaling}) and the
	implicit dependence through the function $f$, with higher-spin
	systems producing systematically higher maximum energies. This
	correlation between gravitational wave observables, specifically the
	inferred remnant spin, and the predicted UHECR production provides a
	testable prediction of this acceleration scenario.
	
	\subsection{BH-NS and BNS channels}
	
	We now comment separately on the BH-NS and BNS channels, whose
	remnant masses and field-sustaining mechanisms differ substantially
	from those of the BBH population discussed above and are not
	represented in Table~\ref{tab:gws}.
	
	For BH-NS binaries with separations on the order of a few neutron-star radii, or for post-merger BHs formed in binary neutron-star mergers, field strengths in the range $B \sim 10^{12}$--$10^{14}$ G have also been suggested. In BH-NS mergers, magnetar companions may possess surface magnetic fields of $10^{14}$--$10^{15}\,\mathrm{G}$, and
	post-merger amplification mechanisms can generate similarly intense
	fields in the surrounding plasma \citep{2010A&A...515A..30O,2015ApJ...809...39G,2015PhRvD..92f4034K,2016ApJ...824L...6R,PhysRevD.106.023013,2020ARNPS..70...95R,2025MNRAS.542.3067A}. We discuss the physical viability of the disk-anchored field
	scenario, and its restriction to the tidal-disruption-favorable mass
	ratio regime, in Sec.~\ref{sec:bns_viability}.
	
	Finally, we note that BNS remnants occupy a mass range,
	$M \sim 2.5$--$4.5\,M_\odot$, well below the $20$--$150\,M_\odot$
	range considered for the BBH population above. No BNS event is
	therefore included in Table~\ref{tab:gws}, and any reference to BNS
	magnetic field strengths elsewhere in this work (e.g.\ Fig.~2)
	should be understood as illustrative of the neutron-star
	magnetospheric field regime, not corresponding to the higher BBH
	remnant masses shown in that figure.

	\section{Results and discussion}
	\label{sec:results}
	
	In this section, we explore in more detail the parameter space of
	center–of–mass collisions around magnetized Kerr black holes in the
	regime relevant for compact binaries that will be observed as GW sources.
	Using the framework described in Sec.~\ref{sec:theory}, we compute the
	center–of–mass energy $E_{\rm cm}$ of proton–proton collisions at a
	radius $r = r_{\rm H} + \varepsilon$ outside the event horizon, for a
	range of magnetic fields $B$, black–hole masses $M$, spin parameters
	$a/M$, and specific angular momenta $\ell_1$ and $\ell_2$ of the
	colliding particles. We stress that UHECR observations tend to favor heavier incoming particles. In our scenario, this corresponds to larger $m_0$, which would simply increase $E_{\rm max}$. Hence, adopting protons as the primary particles should be regarded as a conservative choice.
	
\begin{figure}
		\centering 
		\subfloat[$E_{\rm cm}$ as a function of $B$ for different radial offsets 
		$\varepsilon$ at fixed $M = 10\,M_\odot$, $\chi_f = 0.9$, 
		$\ell_1 = 1.5$ and $\ell_2 = -1.5$.]{
						\includegraphics[width=0.48\textwidth]{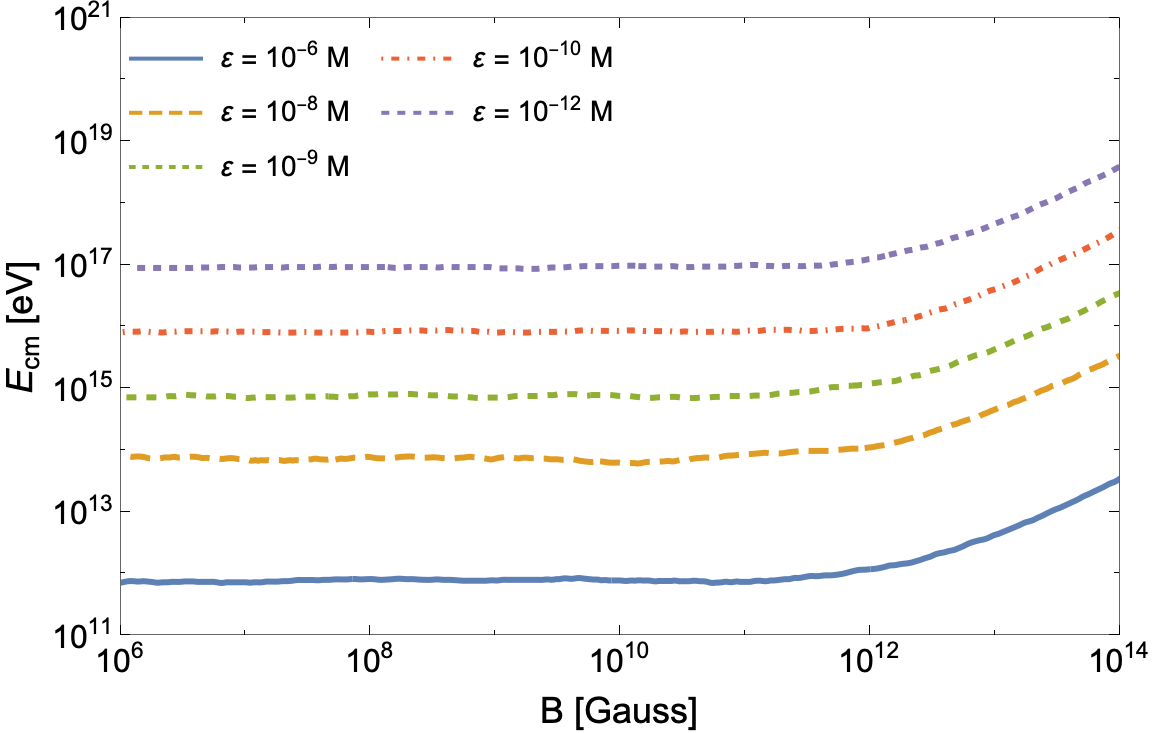}
					}
				\\
		 \subfloat[$E_{\rm cm}$ as a function of $B$ for different 
		 masses at fixed 
		 $\varepsilon = 10^{-10}M$, $\chi_f = 0.9$, 
		 $\ell_1 = 1.5$ and $\ell_2 = -1.5$.]{
			\includegraphics[width=0.48\textwidth]{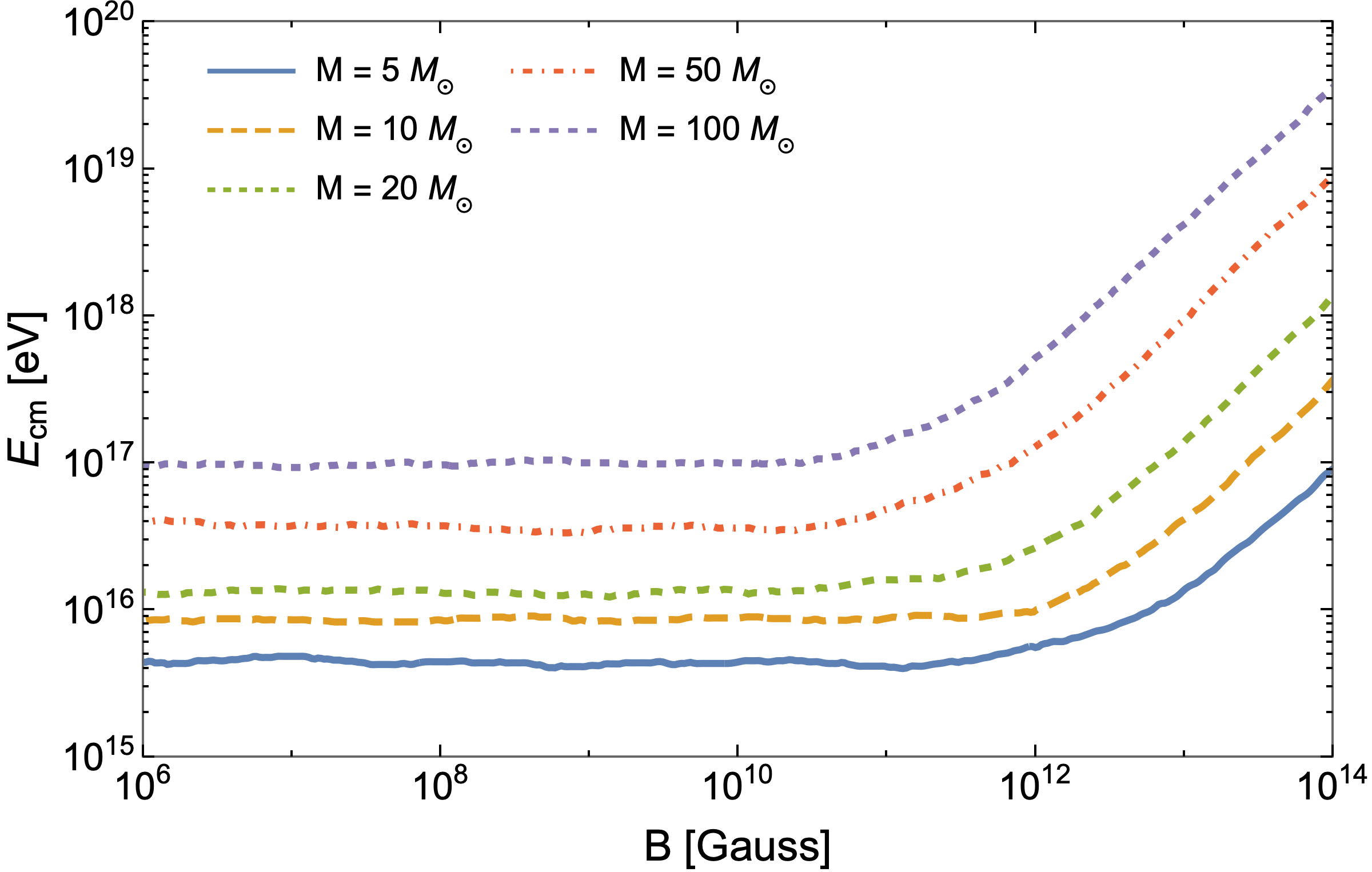}
		}
		\caption{
			\small \justifying Center-of-mass energy $E_{\rm cm}$ of proton--proton collisions near a
			magnetised Kerr black hole as a function of the magnetic field $B$.
			Panel~(a) shows the dependence on the radial offset 
			$\varepsilon$ from the horizon, while panel~(b) shows the dependence on
			the black-hole mass $M$.}
		\label{fig:EversusB}
	\end{figure}

	Figure~\ref{fig:EversusB}-(a) shows $E_{\rm cm}$ as a function of the magnetic field for
	a fixed mass $M = 10\,M_\odot$ and spin $\chi_f = 0.9$, varying the
	radial offset in the interval $\varepsilon = (10^{-6}$ – $10^{-12})M$.
	For weak fields, $B \lesssim 10^{11}\,{\rm G}$, the dependence on
	$\varepsilon$ is modest, with $E_{\rm cm}$ ranging between
	$\sim 10^{12}$ and $10^{16}\,{\rm eV}$ as the collision point is moved
	closer to the horizon.
	For stronger fields, $B \gtrsim 10^{12}\,{\rm G}$, the effect of the
	radial offset becomes more pronounced and the curves separate:
	collisions occurring at $\varepsilon \lesssim 10^{-10}M$ reach
	$E_{\rm cm} \gtrsim 10^{18}\,{\rm eV}$, whereas for
	$\varepsilon \sim 10^{-6}M$ they remain below the ultrahigh–energy
	regime.
	This behaviour confirms that the magnetic BSW mechanism becomes most
	efficient when collisions take place very close to the horizon and in
	the presence of strong magnetic fields, but it also shows that the
	threshold in $\varepsilon$ is not extremely fine–tuned: once
	$\varepsilon \lesssim 10^{-10}M$ the curves nearly saturate and the
	gain in $E_{\rm cm}$ with further inward displacement is relatively
	small.
	
	The apparent bumpiness in Figs.~\ref{fig:EversusB}-(a) and (b) for $B \lesssim 10^{12}$\,G is a sampling artefact of the numerical optimization over $(\ell_1, \ell_2)$. In the gravity-dominated regime, the energy landscape is relatively flat but punctuated by narrow resonant ridges (visible in Fig.~\ref{fig:contoursl}(a)); small changes in $B$ shift these ridges, causing the discrete grid to jump between ridge peaks and the surrounding plateau. For $B \gtrsim 10^{12}$\,G, the magnetic term dominates and the effective potential becomes smoother, yielding correspondingly smooth curves. The envelope of achievable energies is monotonically increasing with $B$.
	
	The impact of the black–hole mass is illustrated in Fig.~\ref{fig:EversusB}-(b), where
	we fix $\varepsilon = 10^{-10}M$, $\chi_f = 0.9$, and the angular momenta
	$\ell_1 = 1.5$ and $\ell_2 = -1.5$, and vary the mass from
	$M = 5$ to $100\,M_\odot$.
	For $B \sim 10^{10}\,{\rm G}$ all masses yield
	$E_{\rm cm} \sim 10^{16}$–$10^{17}\,{\rm eV}$, while for
	$B \sim 10^{14}\,{\rm G}$ the most massive systems reach
	$E_{\rm cm} \sim 10^{19}$–$10^{20}\,{\rm eV}$.
	The nearly parallel curves indicate an approximate scaling
	$E_{\rm cm} \propto M$ at fixed $(B,\varepsilon,\chi_f,\ell_i)$, as
	already suggested by the analytic expression for the form factor
	$\mathcal{F}(\chi_f, \mathcal{B}, \ell_1, \ell_2)$. This implies that the most massive stellar–origin black holes observed in current GW catalogs
	are naturally favored as sources of the highest–energy events, but
	even systems with $M \lesssim 20\,M_\odot$ can produce
	$E_{\rm cm} \gtrsim 10^{18}\,{\rm eV}$ provided that
	$B \gtrsim 10^{13}\,{\rm G}$ and collisions occur sufficiently close
	to the horizon.
	
\begin{figure*}
		\centering 
		\centering
		\includegraphics[width=0.7\textwidth]{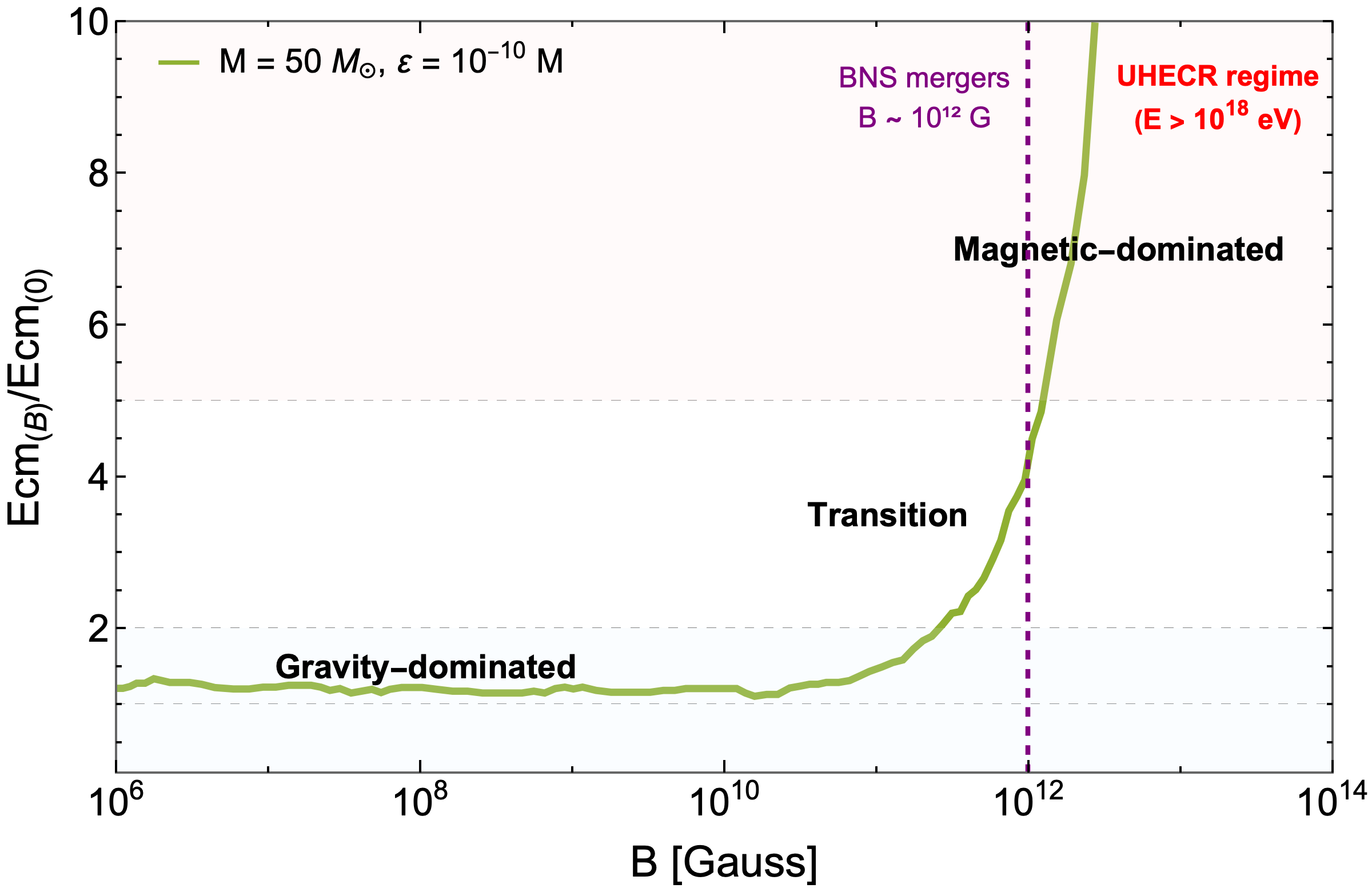}
		\caption{
			\small \justifying Magnetic amplification factor $E_{\rm cm}(B)/E_{\rm cm}(0)$ for proton--proton collisions near a Kerr black hole ($M = 50\,M_{\odot}$, $\chi_f = 0.9$, $\varepsilon = 10^{-10}M$, $\ell_1 = 1.5$, $\ell_2 = -1.5$) as a function of magnetic field strength $B$. The horizontal line at ratio $= 1$ marks the baseline BSW energy (no magnetic enhancement). Three regimes are visible: gravity-dominated ($B < 10^{12}$\,G), transition ($10^{12}\,{\rm G} \lesssim B \lesssim 10^{13}$\,G), and magnetic-dominated ($B > 10^{13}$\,G). The vertical dashed line indicates a fiducial magnetospheric field strength of $B \sim 10^{12}$\,G, relevant for charged BH-BH remnants of comparable mass; this panel does not represent a BNS remnant, whose typical mass ($\sim 2.5$--$4.5\,M_\odot$) lies well outside the range considered here. For $B \gtrsim 10^{13}$\,G, magnetic fields amplify collision energies by nearly an order of magnitude, enabling UHECR production ($E > 10^{18}$\,eV) in astrophysically realistic scenarios.}
		\label{fig:magnetic_ratio}
	\end{figure*}

	To better visualize the relative contribution of the magnetic field to 
	particle acceleration, Fig.~\ref{fig:magnetic_ratio} presents the 
	magnetic amplification factor $E_{\rm cm}(B)/E_{\rm cm}(0)$, which 
	quantifies how many times the presence of a magnetic field enhances 
	the collision energy compared to the purely gravitational BSW mechanism 
	(i.e., $B = 0$). We adopt representative parameters 
	$M = 50\,M_\odot$, $\chi_f = 0.9$, $\varepsilon = 10^{-10}M$, and 
	$\ell_1 = 1.5$, $\ell_2 = -1.5$, which correspond to a near-extremal 
	black hole with collisions occurring very close to the horizon.
	The horizontal dashed line at ratio $= 1$ marks the baseline energy 
	where the magnetic field has no effect, serving as a reference to 
	identify three distinct acceleration regimes.
	
	In the \textit{gravity-dominated regime} ($B < 10^{12}\,{\rm G}$), 
	the amplification factor remains below $\sim 2$, indicating that the 
	magnetic field provides at most a modest enhancement over the purely 
	gravitational mechanism. Here, particle acceleration is driven primarily 
	by the spacetime curvature near the event horizon, with magnetic effects 
	playing a secondary role. This regime is consistent with environments 
	where magnetic fields are relatively weak or the collision geometry 
	does not favor strong Lorentz-force contributions.
	
	As the magnetic field strength increases into the 
	\textit{transition regime} ($10^{12}\,{\rm G} \lesssim B \lesssim 10^{13}\,{\rm G}$), 
	the amplification factor rises steeply from $\sim 2$ to $\sim 5$. 
	In this interval, magnetic and gravitational effects become comparable, 
	and the collision energy becomes increasingly sensitive to the field 
	strength. Notably, this range includes typical magnetic-field values 
	inferred for BH-NS and NS-NS merger environments 
	($B \sim 10^{12}$\,G, marked by the purple vertical line), where 
	post-merger outflows and magnetospheric reconnection can naturally 
	generate strong fields \citep[e.g.,][]{2025PhRvL.134h1003F, 2025ApJ...994L...7F}. The fact that 
	realistic NS field strengths fall squarely within this transition 
	regime suggests that magnetic enhancement of BSW collisions may be a 
	generic feature of mergers involving NSs, rather than an exotic or fine-tuned 
	scenario.
	
	For $B > 10^{13}\,{\rm G}$, the system enters the 
	\textit{magnetic-dominated regime}, where the amplification factor 
	exceeds $\sim 5$ and approaches $\sim 10$ at $B \sim 10^{14}$\,G indicating magnetar-strength fields. In this 
	regime, the magnetic field overwhelms the gravitational contribution 
	and becomes the dominant driver of particle acceleration. The tenfold 
	enhancement implies that collisions which would yield 
	$E_{\rm cm} \sim 10^{19}\,{\rm eV}$ in the absence of a magnetic field 
	can reach $E_{\rm cm} \sim 10^{20}\,{\rm eV}$ when $B \sim 10^{14}$\,G. 
	This places the collision energies well into the UHECR regime. While magnetar-strength fields are less common 
	than typical BNS and BH-NS merger fields, they may occur transiently during the 
	most violent phases of compact binary coalescence or in the immediate 
	vicinity of newly formed hypermassive neutron stars 
	\citep[e.g.,][]{2015ApJ...809...39G, 2020GReGr..52...59C}.
	
	The three-regime structure revealed in Fig.~\ref{fig:magnetic_ratio} 
	has important implications for identifying astrophysical sources of 
	UHECRs. First, it demonstrates that magnetic fields do not need to 
	reach extreme values to significantly enhance particle acceleration: 
	even moderate fields of $B \sim 10^{12}$--$10^{13}$\,G, which are 
	well within the range of current numerical simulations of BNS mergers, 
	can double or triple the collision energy. Second, the rapid transition 
	from the gravity-dominated to the magnetic-dominated regime occurs 
	over a relatively narrow range in $\log_{10}(B)$, suggesting that 
	observations of UHECRs correlated with gravitational-wave events could 
	place strong constraints on the magnetic-field strengths in these 
	systems. Finally, the figure underscores the robustness of the magnetic 
	BSW mechanism: unlike the original BSW scenario, which required 
	extremely fine-tuned parameters to achieve unbounded energies, the 
	presence of even moderately strong magnetic fields enables UHECR 
	production across a broad range of astrophysically realistic 
	configurations. It is important to emphasize that $B\sim 10^{12}$~G does not redefine the threshold for ultra-high-energy particle production obtained in our previous work \cite{2024PhRvL.132i1401P}. Fields of order $10^{10}-10^{12}\,\mathrm{G}$ remain sufficient to generate UHECR-scale collision energies. The transition identified here instead marks the point at which magnetic effects become the dominant contribution to the center-of-mass energy, providing a refined physical interpretation of the acceleration mechanism.
	
\begin{figure*} 
		\subfloat[Contour map of $\log_{10}(E_{\rm cm})$ as a function of 
		$\log_{10}(B)$ and $\log_{10}(M/M_\odot)$ for particle collisions.]{
			\includegraphics[width=0.48\textwidth]{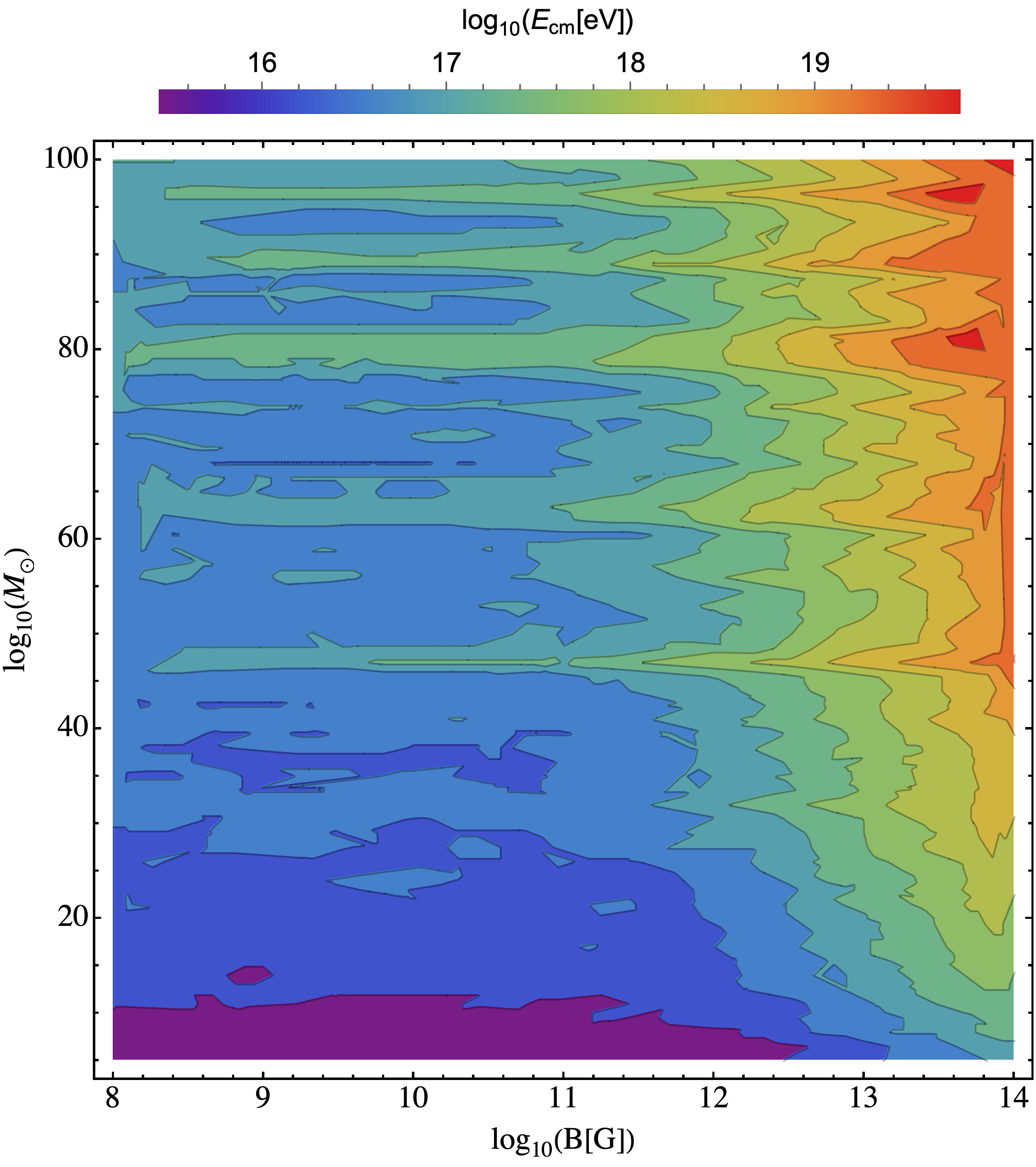}
		}
		\subfloat[Contour map of $\log_{10}(E_{\rm cm})$ as a function of 
		$\log_{10}(B)$ and the spin parameter $\chi_f$ for the same setup.]{
			\includegraphics[width=0.48\textwidth]{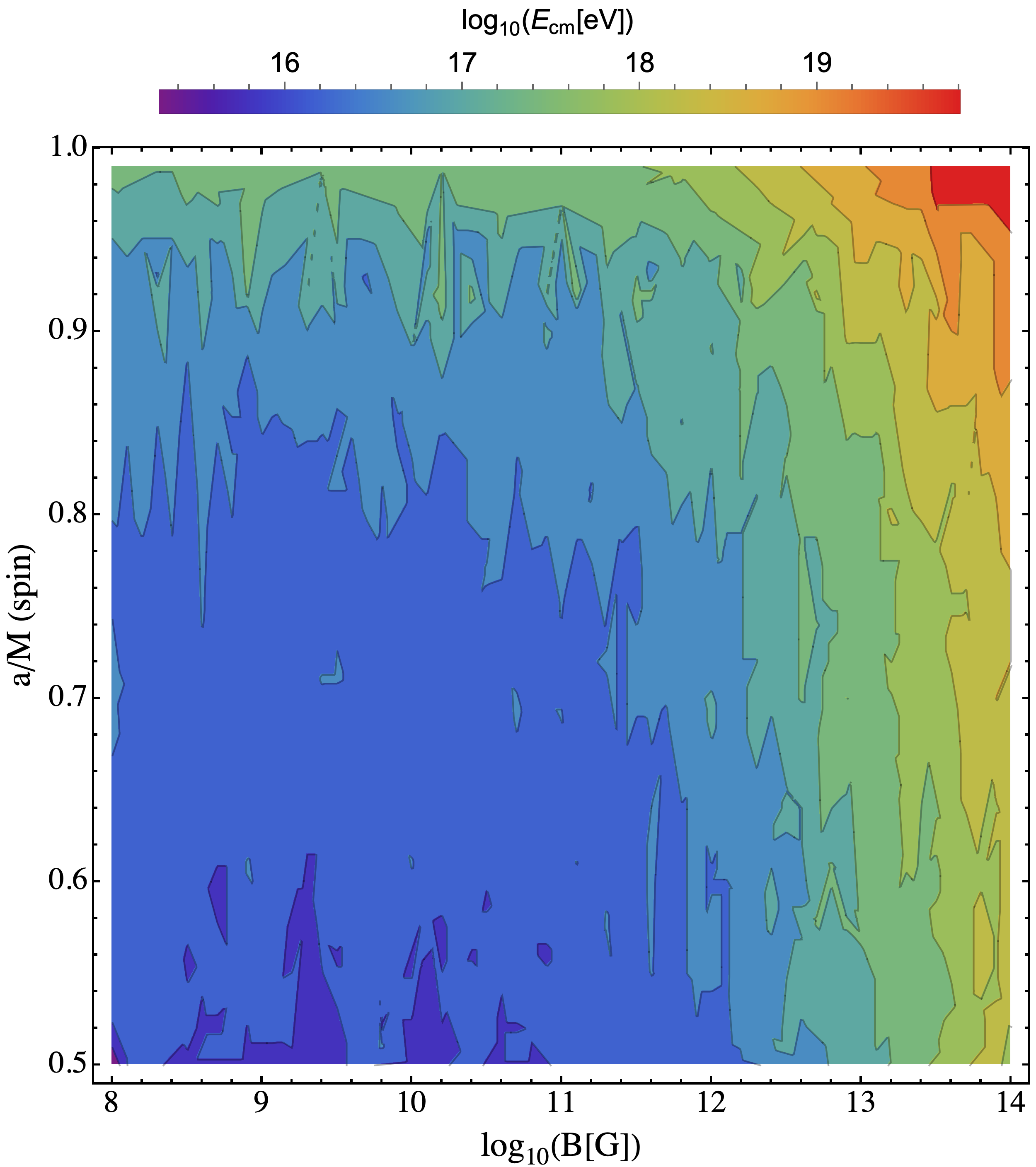}
		}
		\caption{
			\small \justifying Contour maps of the center-of-mass energy $E_{\rm cm}$ for proton--proton
			collisions close to the horizon of a magnetised Kerr black hole, in a regime
			relevant for stellar-mass black holes involved in gravitational-wave mergers.
			In both panels we adopt a fixed radial offset $r = r_{\rm H} + \varepsilon$
			with $\varepsilon = 10^{-10}M$ and specific angular momenta 
			$\ell_1 = 1.5$ and $\ell_2 = -1.5$.  
			Colours indicate $\log_{10}(E_{\rm cm}\,[\mathrm{eV}])$. 
			Panel~(a) shows the dependence on magnetic-field strength and black-hole mass
			for spin $\chi_f = 0.9$, whereas panel~(b) shows the dependence on magnetic-field
			strength and spin for a fixed mass $M = 50\,M_\odot$.}
		\label{fig:contoursB}
	\end{figure*}

\begin{figure*}
		\subfloat[$B = 10^{12}$ {G}]{
			\includegraphics[width=0.48\textwidth]{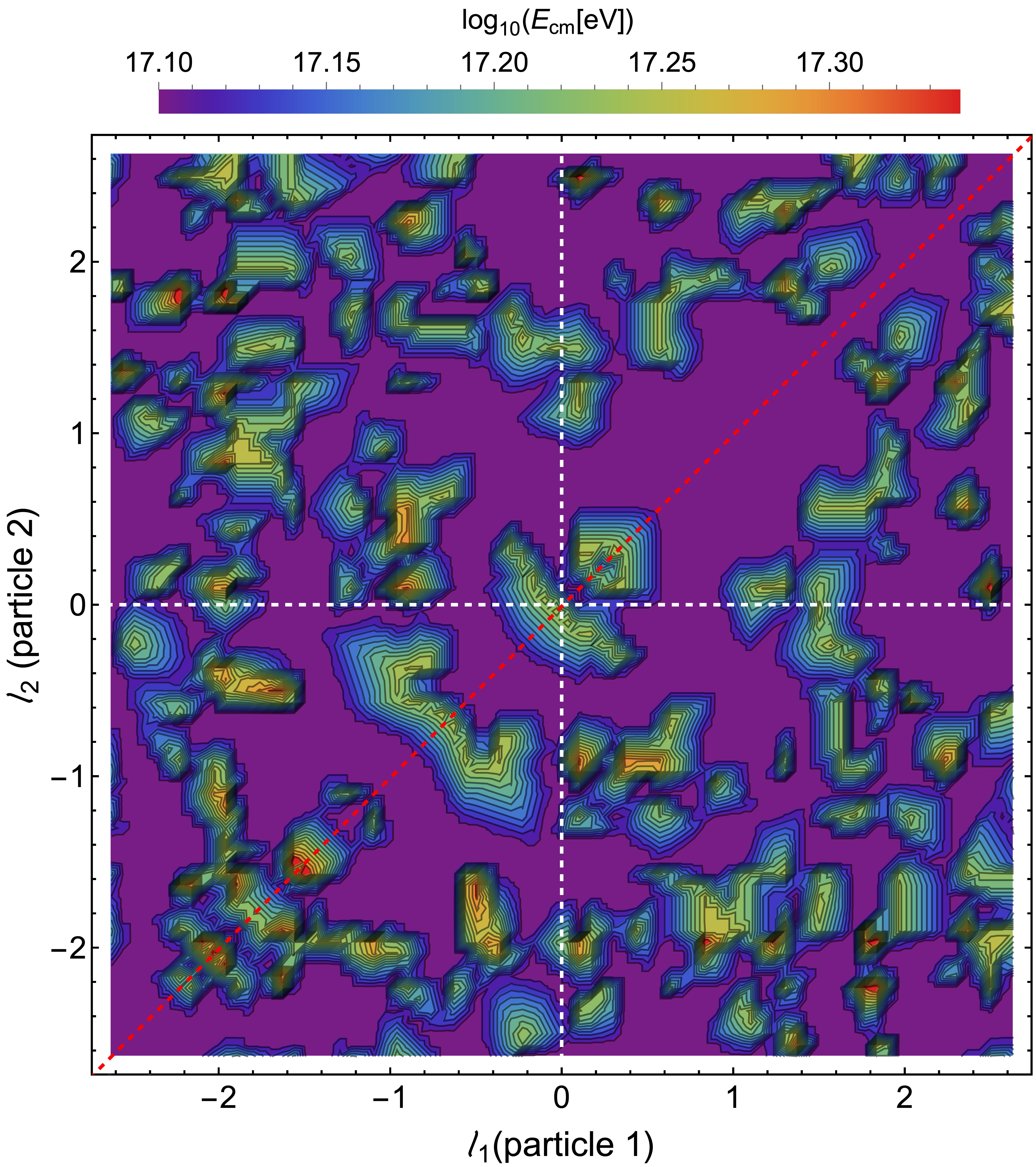}
		}
		\subfloat[$B = 10^{14}$ {G}]{		
			\includegraphics[width=0.48\textwidth]{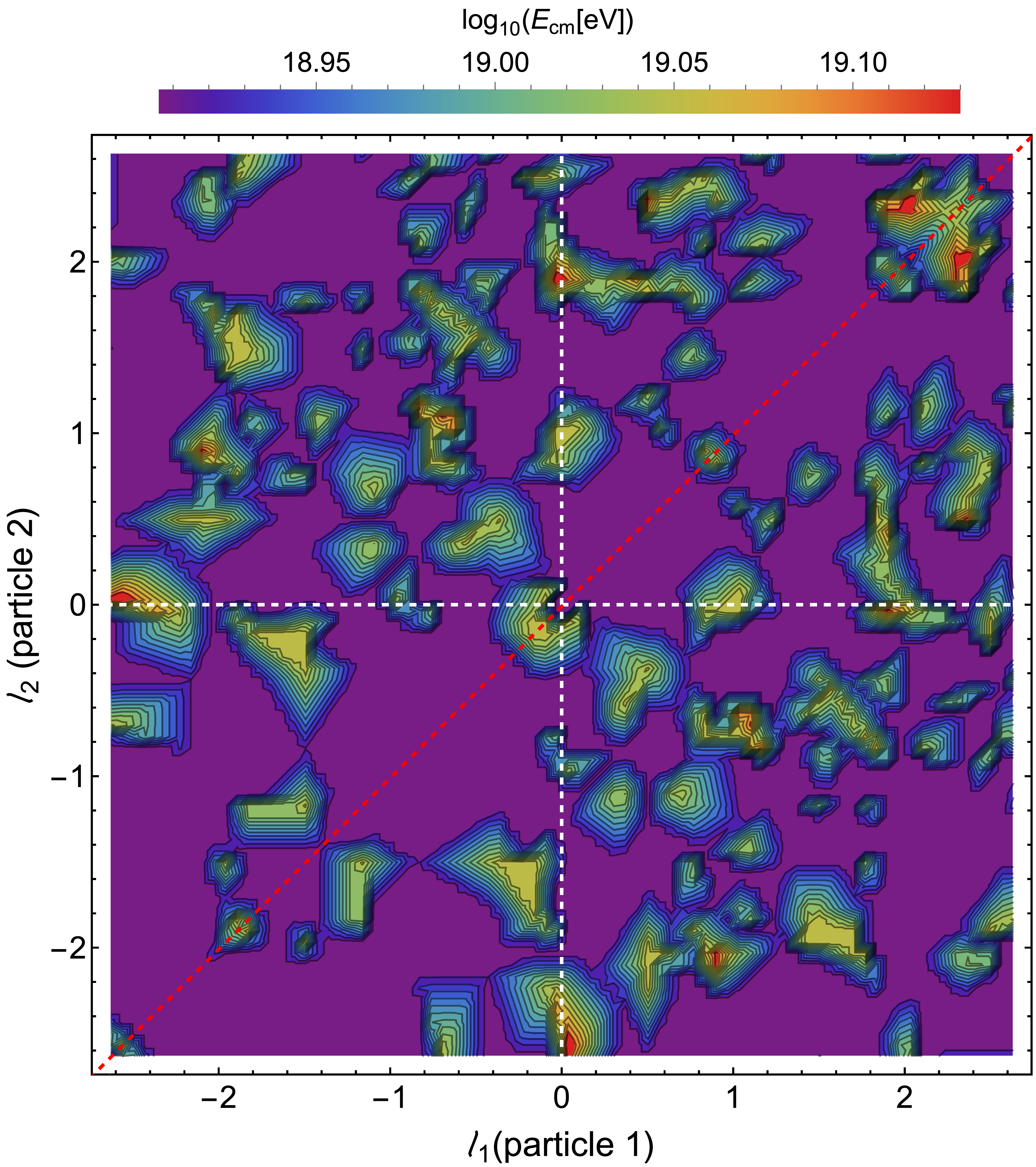}
		}
		\caption{\small \justifying Contour map of $\log_{10}(E_{\rm cm}\,[\mathrm{eV}])$ as a function of the
			specific angular momenta of the two colliding particles, $\ell_1$ (particle 1)
			and $\ell_2$ (particle 2). The collisions occur near the horizon at
			$r = r_{\rm H} + \varepsilon$ with $\varepsilon = 10^{-10}M$ for a magnetised
			Kerr black hole of mass $M = 50\,M_\odot$, spin parameter $\chi_f = 0.9$, and
			magnetic field $B = 10^{12}\,\mathrm{G}$. White dashed lines indicate
			$\ell_1 = 0$ and $\ell_2 = 0$, while the red dashed diagonal marks
			$\ell_1 = \ell_2$, along which the configuration is symmetric under the
			exchange of the two particles.}
		\label{fig:contoursl}
	\end{figure*}

	To visualize the global structure of the parameter space, in Fig.~\ref{fig:contoursB} we
	present contour maps of $\log_{10}(E_{\rm cm}/{\rm eV})$. In both panels, we adopt the near–horizon radius
	$r = r_{\rm H} + \varepsilon$ with $\varepsilon = 10^{-10}M$ and the
	same angular momenta $\ell_1 = 1.5$ and $\ell_2 = -1.5$.
	Figure~\ref{fig:contoursB}-(a) spans the plane $(\log_{10}B,\log_{10}M)$ at fixed
	spin $\chi_f = 0.9$.
	The contours are almost vertical for $B \lesssim 10^{11}\,{\rm G}$,
	indicating a weak dependence on $B$ and a mild increase of
	$E_{\rm cm}$ with mass.
	At $B \gtrsim 10^{12}\,{\rm G}$ the contours bend and $E_{\rm cm}$
	rises steeply, surpassing $10^{18}\,{\rm eV}$ for
	$M \gtrsim 20\,M_\odot$ and approaching $10^{20}\,{\rm eV}$ for
	$M \sim 100\,M_\odot$ and $B \sim 10^{14}\,{\rm G}$.
	This plot shows a broad region of parameter space in which
	ordinary neutron–star magnetospheric fields or fields sourced by small
	charge–to–mass ratios of the black holes are sufficient to drive
	ultrahigh–energy collisions in systems compatible with current GW
	observations.
	
	The horizontal band structures in Fig.~\ref{fig:contoursB}-(a) have the same origin as the bumpiness in Figs.~\ref{fig:EversusB}, but they manifest along the mass axis. As $M$ varies, the allowed $(\ell_1,\ell_2)$ range shifts, causing the resonant ridges in angular-momentum space to drift. At certain masses, the optimal configuration aligns with a critical trajectory near the horizon, temporarily elevating $E_{\rm cm}$ above the smooth trend. Between these masses, the grid falls between ridges and records a lower value. These deviations are expected artifacts of the finite angular-momentum grid.
	
	Figure~\ref{fig:contoursB}-(b) explores instead the $(\log_{10}B,\chi_f)$ plane for a fixed
	mass $M = 50\,M_\odot$.
	For $B \lesssim 10^{11}\,{\rm G}$ the dependence on the spin parameter
	is weak: $E_{\rm cm}$ varies by less than an order of magnitude when
	$\chi_f$ is changed from $0.5$ to nearly extremal values, remaining below
	$10^{18}\,{\rm eV}$.
	In contrast, when $B \gtrsim 10^{12}\,{\rm G}$ the contours become
	more tilted, and both higher spin and stronger magnetic fields
	contribute to increasing $E_{\rm cm}$.
	Ultrahigh–energy collisions appear already for
	$\chi_f \simeq 0.6$–$0.7$ at $B \sim 10^{13}\,{\rm G}$ and extend up to
	$E_{\rm cm} \gtrsim 10^{19}\,{\rm eV}$ for nearly extremal spins and
	$B \sim 10^{14}\,{\rm G}$.
	Compared with the original BSW scenario, which required
	fine–tuning to $\chi_f \to 1$, the presence of a magnetic field therefore
	dramatically enlarges the allowed range of spin parameters:
	for realistic $B$, the production of UHE events is possible for almost
	the entire distribution of spins inferred from GW catalogs.
	
	Finally, Fig.~\ref{fig:contoursl} presents contour maps of $\log_{10}(E_{\rm cm}/{\rm eV})$
	in the $(\ell_1,\ell_2)$ plane for a magnetised Kerr black hole with
	$M = 50\,M_\odot$, $\chi_f = 0.9$ and $\varepsilon = 10^{-10}M$, for two
	representative magnetic-field strengths, $B = 10^{12}\,{\rm G}$ [panel
	(a)] and $B = 10^{14}\,{\rm G}$ [panel (b)]. In both cases we explore
	$\ell_i \in [-2.63,2.63]$, and the maps exhibit a patchy pattern of narrow
	ridges and islands, corresponding to local maxima of $E_{\rm cm}$ where
	the particle orbits approach critical trajectories. The red dashed
	diagonal marks $\ell_1 = \ell_2$, along which the configuration is
	symmetric under the exchange of the two particles, while the white dashed
	lines indicate $\ell_1 = 0$ and $\ell_2 = 0$. For $B = 10^{12}\,{\rm G}$
	the dynamic range is relatively modest, with
	$\log_{10}(E_{\rm cm}/{\rm eV}) \simeq 17.1$–$17.3$ across most of the
	plane, and broad regions around $\ell_1 \approx 0$ and $\ell_2 \approx 0$
	already yielding $E_{\rm cm} \gtrsim 10^{17}\,{\rm eV}$. Increasing the
	field to $B = 10^{14}\,{\rm G}$ shifts the entire pattern upward by about
	two orders of magnitude so that typical configurations reach
	$E_{\rm cm} \sim 10^{19}\,{\rm eV}$, and the highest ridges slightly
	exceed this value. These maps show that, once the magnetic field, mass,
	spin and radial offset are chosen in the UHE–favourable regime,
	ultrahigh–energy collisions do not rely on extremely fine–tuned angular
	momenta: while the very highest energies are associated with special
	combinations of $(\ell_1,\ell_2)$, a broad portion of phase space
	produces $E_{\rm cm}$ within an order of magnitude of the extrema.
	
	The island structure visible in Fig.~\ref{fig:contoursl} reflects the genuine topology of the BSW energy functional. The $E_{\rm cm}$ exhibits sharp, localized maxima near critical angular-momentum values ($\ell_{\rm crit} \approx 2$) where particles approach the horizon asymptotically, departing from these values causes a rapid energy drop. The feasibility constraint $\dot{r}^2 \geq 0$ (Eq.~(\ref{eq:rdot})) further excludes angular-momentum combinations corresponding to trapped trajectories, carving out disconnected viable regions. The magnetic field introduces additional families of critical orbits via its $B$-dependent shift of the effective potential, redistributing the island pattern, as seen comparing panels (a) and (b). The absence of smooth variation thus reflects the highly non-linear, multi-peaked structure of $E_{\rm cm}(\ell_1,\ell_2)$ in Kerr spacetime.
	
	To connect our theoretical framework with actual gravitational wave observations, we apply our calculations to the complete catalog of high-spin merger remnants detected by LIGO-Virgo-KAGRA. Table~\ref{tab:gws} lists all 34 gravitational wave events from the GWTC-2.1, GWTC-3, and GWTC-4.0 catalogs with final spin parameters exceeding $\chi_f > 0.7$~\cite{gwtc2,gwtc3,abbott2023population}. For each event, we report the redshift $z$, the component masses before merger ($m_1$ and $m_2$), the final remnant mass $M$, the final dimensionless spin $\chi_f$, and our computed maximum center-of-mass energy $\log_{10}(E_{\rm max}$ [eV]) assuming collisions at $\varepsilon = 10^{-10}M$ from the horizon with magnetic field $B = 10^{14}$ G and optimized angular momenta within the physically allowed range given by Eq.~\ref{eq:ell_range}.
	
	The predicted maximum energies span approximately three orders of magnitude, from $\log_{10}(E_{\rm max}) \approx 17.8$ for the lowest-mass, lowest-spin systems to $\log_{10}(E_{\rm max}) \approx 20.2$ for the most massive, highest-spin events. The highest predicted energy corresponds to GW200308\_173609, with a final mass $M = 88^{+169}_{-47}\,M_\odot$ and spin $\chi_f = 0.91^{+0.03}_{-0.08}$, yielding $E_{\rm max} \sim 1.6 \times 10^{20}$ eV. This places it at the extreme end of the observed UHECR spectrum. Other notable high-energy events include GW190519\_153544 ($E_{\rm max} \sim 10^{20}$ eV, $M = 100.0^{+13.0}_{-12.9}\,M_\odot$, $\chi_f = 0.79^{+0.07}_{-0.13}$) and GW190706\_222641 ($E_{\rm max} \sim 7.9 \times 10^{19}$ eV, $M = 107.3^{+25.2}_{-15.9}\,M_\odot$, $\chi_f = 0.78^{+0.09}_{-0.18}$).
	
	The distribution of predicted energies reflects both the mass and spin distributions of the observed merger population. Events with final masses below $\sim 30\,M_\odot$ typically produce maximum energies below the UHECR threshold of $10^{18}$ eV even at our fiducial field strength, while systems with $M \gtrsim 50\,M_\odot$ and $\chi_f \gtrsim 0.75$ consistently achieve energies in the UHECR regime. The approximate linear scaling $E_{\rm max} \propto M$ evident in Fig.~\ref{fig:EversusB}-(b) is clearly visible in the catalog: comparing events with similar spins but different masses, such as GW151226 ($M \approx 21\,M_\odot$, $\log_{10} E_{\rm max} = 18.05$) and GW170729 ($M \approx 80\,M_\odot$, $\log_{10} E_{\rm max} = 19.56$), shows that a factor of $\sim 4$ increase in mass produces approximately a factor of $\sim 30$ increase in maximum energy, consistent with the mass scaling and the implicit spin dependence through the amplification function $\mathcal{F}$.
	
	Notably, the sole binary neutron star event in our catalog, GW170817, has upper limits of $M \leq 2.8\,M_\odot$ and $\chi_f \leq 0.89$ that would place it well below the UHECR regime for our assumed magnetic field strength. However, BNS mergers may feature substantially different magnetic field configurations and particle injection mechanisms compared to BBH systems, making direct comparison problematic. The absence of confirmed BH-NS mergers with $\chi_f > 0.7$ in the current catalog prevents us from making specific predictions for this particularly promising class of UHECR sources, though our general framework applies equally well once such systems are detected and characterized.
	
	Taken together, these results reinforce and extend the picture
	presented in our previous Letter~\cite{2024PhRvL.132i1401P}: magnetised compact binaries are
	natural sites for accelerating particles to ultrahigh energies via the
	magnetic BSW mechanism. For typical parameters of charged BH-BH systems with masses $M \sim 10$--$100\,M_\odot$, or BH-NS systems in the
	disruption-favorable mass ratio regime ($M_{\rm BH} \lesssim 7\,M_\odot$), magnetic fields in the range $B \sim 10^{11}$--$10^{13}$\,G are sufficient to produce $E_{\rm cm} \gtrsim 10^{18}$\,eV over a wide range of spins and particle angular momenta. This robustness with respect to source parameters suggests that UHECR
	production in compact–binary mergers could be a generic phenomenon
	rather than a rare, fine–tuned occurrence, and motivates detailed
	studies of particle escape and propagation, as well as joint
	UHECR–GW searches, to further test this scenario.

	\subsection{Astrophysical viability of the magnetic BSW mechanism in BNS post-merger environments}
	\label{sec:bns_viability}
	
	A natural extension of this work is the post-merger production of UHECRs in binary neutron-star coalescences. Unlike the pre-merger phase, this channel relies on the formation of a metastable, often differentially rotating and strongly magnetized neutron-star remnant, or its subsequent collapse into a black hole \cite{2017ApJ...850L..19M, 2019ARNPS..69...41S}. 
	We acknowledge that a passive vacuum dipole decay ($r^{-3}$) from a companion would require non-physical surface fields for high-mass systems. However, in realistic merger and post-merger environments, the magnetic flux is ``anchored'' by high-conductivity plasma, baryon-loaded ejecta, and accretion debris \cite{2012PhRvD..85b4020F,2010PhRvD..82h3004Z}. This flux anchoring mechanism maintains high field densities in the near-horizon region independently of the companion's initial dipole distance, providing the necessary physical framework to sustain the required field strengths. Whether sourced by flux inherited from the remnant or amplified by disk-driven MHD processes \cite{PhysRevD.97.124039}, the black hole's ergosphere remains a viable site for particle acceleration. We stress that the disk-anchored field argument above applies only
	to BH-NS systems in the disruption-favorable regime,
	$q \equiv M_{\rm BH}/M_{\rm NS} \lesssim 5$, i.e.\
	$M_{\rm BH} \lesssim 7\,M_\odot$ for a $1.4\,M_\odot$ companion, even
	at near-extremal spin \cite{2021PhRvD.103f4007F}. For the more massive
	remnants populating Table~\ref{tab:gws}
	($M \sim 20$--$150\,M_\odot$), the mass ratio is far too large for
	tidal disruption to occur, no disk forms, and the relevant
	field-sustaining mechanism there is instead the charge-to-mass ratio
	scenario for BH-BH systems described above, or fields inherited
	directly from residual accretion in the pre-merger phase. We
	therefore do not invoke post-merger disk formation for the BBH
	population analyzed in Table~\ref{tab:gws}.
	
	The availability of charged particles is ensured by baryon-loaded ejecta, fallback material, and disk/wind outflows, which provide a copious plasma reservoir for injection into the ergosphere. While transient, these magnetic configurations may persist long enough to be astrophysically significant, depending on the survival time of the accretion disk and reconnection timescales \cite{2014A&A...562A.137F}. Furthermore, if the collapse to a black hole is delayed, magnetic-field amplification mechanisms operating in the hypermassive neutron-star remnant can significantly increase the field strength prior to collapse. Kelvin–Helmholtz instabilities generated during merger and magnetorotational-instability-driven turbulence in the remnant and accretion flow are known to amplify magnetic fields by several orders of magnitude \citep{2010A&A...515A..30O,2015ApJ...809...39G,2015PhRvD..92f4034K,2016ApJ...824L...6R}. GRMHD simulations have reported magnetic-field strengths reaching magnetar levels ($\sim10^{15}\,\mathrm{G}$) and, in some regions, even larger values \citep{2016ApJ...824L...6R,PhysRevD.106.023013,2025MNRAS.542.3067A}. The newly formed black hole may therefore be embedded in an environment considerably more magnetized than the original neutron stars. This motivates a dual-stage (pre- and post-merger) acceleration framework that would be falsifiable via multimessenger observations \cite{2017ApJ...848L..12A}. 
	
	The background metric can be treated as effectively stationary throughout the process. For a $150 M_{\odot}$ black hole, the light-crossing times are on the millisecond scale, which is perfectly compatible with the ultra-relativistic interaction and escape timescales of the BSW mechanism. There is a clear hierarchy of timescales: the spacetime settles into a stationary state over the damping time ($\tau_{QNM} \approx 0.15$~ms for $3 M_{\odot}$, scaled proportionally for higher masses) \cite{Leaver1985,Berti2009}, while the magnetic configuration anchored by the disk persists for $\tau_{disk} \sim 0.1$ to $1$~s \cite{2019ARNPS..69...41S}. Since $\tau_{disk} \gg \tau_{QNM}$, the ``accelerator'' remains active and stable for a duration orders of magnitude longer than the final plunge dynamics, providing a robust physical environment for UHECR production even in high-mass remnants where linear mass scaling ($E_{max} \propto M$) compensates for more conservative field strengths. Further analysis of the above points is left for future work.

	\newpage
	
	\begin{table*}
		\centering
		\renewcommand{\arraystretch}{1}
		\caption{\small \justifying List of LIGO-Virgo-KAGRA sources with rotational parameter $\chi_f\equiv a/M\equiv J/M^2 > 0.7$ ($J$ is the BH's angular momentum) for the BH produced after the merger of BBHs/BH-NS/BNS. Updated with latest GWTC-2.1, GWTC-3, and GWTC-4.0 measurements. The list is ordered following the chronological detections. The masses of the BHs/NSs before the merger are denoted by $m_1$ and $m_2$, and $M$ is the mass of the final BH/NS. All masses are in solar mass (M$_\odot$) units. The redshift of a source is $z$. $\log_{10} E$ is the maximum center-of-mass energy (in eV) for particle collisions at $\varepsilon=10^{-10}M$ from the horizon, calculated using the BSW mechanism with magnetic field $B=10^{14}$ G and optimal angular momenta within the allowed range $-2(1+\sqrt{1-\chi_f}) < \ell < 2(1+\sqrt{1-\chi_f})$. The events are classified here as binary of black holes (BBHs) or binary of black hole-neutron star (BH-NS) or binary of neutron stars (BNSs).}
		\label{tab:gws}
		\begin{tabular}{lllllllll}
			\hline
			\hline
			Catalog & Event       & $z$ & $m_1$ & $m_2$  & $M$      & $\chi_f$  & $\log_{10}E$ & Type       \\ \hline
			GWTC-2.1 & GW151226    & 0.10$^{+0.03}_{-0.04}$  & 14.2$^{+11.1}_{-3.6}$ & 7.5$^{+2.4}_{-2.8}$ & 20.7$^{+8.6}_{-1.6}$ & 0.74$^{+0.07}_{-0.05}$ & 18.05 & BBH \\
			GWTC-2.1 & GW170729    & 0.44$^{+0.24}_{-0.19}$ & 54.7$^{+12.7}_{-12.8}$ & 30.2$^{+11.9}_{-10.2}$ & 80.3$^{+13.5}_{-10.2}$ & 0.81$^{+0.07}_{-0.13}$ & 19.56 & BBH\\ 
			GWTC-2.1 & GW170814 & 0.13$^{+0.03}_{-0.05}$ & 30.9$^{+5.4}_{-3.3}$ & 24.9$^{+3.0}_{-4.0}$ & 53.2$^{+3.2}_{-2.7}$ & 0.72$^{+0.07}_{-0.05}$ & 18.91 & BBH\\
			GWTC-2.1 & GW170817$^*$  & 0.01$^{+0.00}_{-0.00}$ & 1.46$^{+0.12}_{-0.10}$ & 1.27$^{+0.09}_{-0.09}$ & $\leq 2.8$ & $\leq 0.89$ & -- & BNS\\
			GWTC-2.1 & GW170823    & 0.36$^{+0.13}_{-0.15}$ & 38.3$^{+9.5}_{-6.2}$ & 29.0$^{+6.5}_{-7.8}$ & 63.9$^{+9.6}_{-6.8}$ & 0.72$^{+0.09}_{-0.12}$ & 19.06 & BBH\\ 
			GWTC-2.1 & GW190517$\_$055101 & 0.33$^{+0.26}_{-0.15}$ & 39.2$^{+13.9}_{-9.2}$ & 24.0$^{+7.4}_{-7.9}$ & 60.1$^{+9.9}_{-9.4}$ & 0.87$^{+0.05}_{-0.07}$ & 19.14 & BBH\\
			GWTC-2.1 & GW190519$\_$153544   & 0.45$^{+0.24}_{-0.15}$  & 65.1$^{+10.8}_{-11.0}$ & 40.8$^{+11.5}_{-12.7}$ & 100.0$^{+13.0}_{-12.9}$ & 0.79$^{+0.07}_{-0.13}$ & 20.05 & BBH \\
			GWTC-2.1 & GW190521  & 0.56$^{+0.36}_{-0.27}$ & 98.4$^{+33.6}_{-21.7}$ & 57.2$^{+27.1}_{-30.1}$ & 147.4$^{+40.0}_{-16.0}$ & 0.71$^{+0.12}_{-0.16}$ & 19.51 & BBH\\ 
			GWTC-2.1 & GW190521$\_$074359 & 0.21$^{+0.10}_{-0.10}$ & 43.4$^{+5.8}_{-5.5}$ & 33.4$^{+5.2}_{-6.8}$ & 72.6$^{+6.5}_{-5.4}$ & 0.72$^{+0.05}_{-0.07}$ & 19.26 & BBH\\
			GWTC-2.1 & GW190527$\_$092055    & 0.44$^{+0.29}_{-0.19}$ & 35.6$^{+18.7}_{-8.0}$ & 22.2$^{+9.0}_{-8.7}$ & 55.5$^{+17.9}_{-8.5}$ & 0.71$^{+0.12}_{-0.16}$ & 18.91 & BBH\\
			GWTC-2.1 & GW190620$\_$030421    & 0.50$^{+0.23}_{-0.20}$ & 58.0$^{+19.2}_{-13.3}$ & 35.0$^{+13.1}_{-14.5}$ & 88.0$^{+17.2}_{-12.4}$ & 0.79$^{+0.08}_{-0.15}$ & 19.36 & BBH\\ 
			GWTC-2.1 & GW190706$\_$222641 & 0.60$^{+0.33}_{-0.29}$ & 74.0$^{+20.1}_{-16.9}$ & 39.4$^{+18.4}_{-15.4}$ & 107.3$^{+25.2}_{-15.9}$ & 0.78$^{+0.09}_{-0.18}$ & 19.90 & BBH\\
			GWTC-2.1 & GW190719$\_$215514 & 0.61$^{+0.39}_{-0.30}$ & 36.6$^{+42.1}_{-11.1}$ & 19.9$^{+10.0}_{-9.3}$ & 54.5$^{+38.3}_{-11.1}$ & 0.78$^{+0.11}_{-0.17}$ & 18.84 & BBH\\
			GWTC-2.1 & GW190720$\_$000836    & 0.16$^{+0.11}_{-0.05}$ & 14.2$^{+5.6}_{-3.3}$ & 7.5$^{+2.2}_{-1.8}$ & 20.8$^{+3.9}_{-2.0}$ & 0.72$^{+0.06}_{-0.05}$ & 17.94 & BBH\\ 
			GWTC-2.1 & GW190727$\_$060333 & 0.52$^{+0.18}_{-0.18}$ & 38.9$^{+8.9}_{-6.0}$ & 30.2$^{+6.5}_{-8.3}$ & 65.4$^{+9.5}_{-7.3}$ & 0.73$^{+0.10}_{-0.10}$ & 18.91 & BBH\\
			GWTC-2.1 & GW190728$\_$064510 & 0.18$^{+0.05}_{-0.07}$ & 12.5$^{+6.9}_{-2.3}$ & 8.0$^{+1.7}_{-2.6}$ & 19.7$^{+4.4}_{-1.4}$ & 0.71$^{+0.04}_{-0.04}$ & 17.79 & BBH\\
			GWTC-2.1 & GW190828$\_$063405 & 0.38$^{+0.10}_{-0.15}$ & 31.9$^{+5.4}_{-4.1}$ & 25.8$^{+4.9}_{-5.3}$ & 54.3$^{+7.3}_{-4.0}$ & 0.75$^{+0.06}_{-0.07}$ & 18.91 & BBH\\
			GWTC-2.1 & GW190930$\_$133541 & 0.16$^{+0.06}_{-0.06}$ & 14.2$^{+8.0}_{-4.0}$ & 6.9$^{+2.4}_{-2.1}$ & 20.2$^{+6.1}_{-2.0}$ & 0.72$^{+0.07}_{-0.06}$ & 17.79 & BBH\\
			GWTC-3 & GW191103$\_$012549    & 0.20$^{+0.09}_{-0.09}$ & 11.8$^{+6.2}_{-2.2}$ & 7.9$^{+1.7}_{-2.4}$ & 19.0$^{+3.8}_{-1.7}$ & 0.75$^{+0.06}_{-0.05}$ & 17.94 & BBH\\ 
			GWTC-3 & GW191126$\_$115259 & 0.30$^{+0.12}_{-0.13}$ & 12.1$^{+5.5}_{-2.2}$ & 8.3$^{+1.9}_{-2.4}$ & 19.6$^{+3.5}_{-2.0}$ & 0.75$^{+0.06}_{-0.05}$ & 17.94 & BBH\\
			GWTC-3 & GW191127$\_$050227 & 0.57$^{+0.40}_{-0.29}$ & 53$^{+47}_{-20}$ & 24$^{+17}_{-14}$ & 76$^{+39}_{-21}$ & 0.75$^{+0.13}_{-0.29}$ & 19.45 & BBH\\
			GWTC-3 & GW191204$\_$110529 & 0.34$^{+0.25}_{-0.18}$ & 27.3$^{+11.0}_{-6.0}$ & 19.3$^{+5.6}_{-6.0}$ & 45.0$^{+8.6}_{-7.6}$ & 0.71$^{+0.12}_{-0.11}$ & 18.54 & BBH\\
			GWTC-3 & GW191204$\_$171526 & 0.13$^{+0.04}_{-0.05}$ & 11.9$^{+3.3}_{-1.8}$ & 8.2$^{+1.4}_{-1.6}$ & 19.21$^{+1.79}_{-0.95}$ & 0.73$^{+0.03}_{-0.03}$ & 17.79 & BBH\\
			GWTC-3 & GW200112$\_$155838 & 0.24$^{+0.07}_{-0.08}$ & 35.6$^{+6.7}_{-4.5}$ & 28.3$^{+4.4}_{-5.9}$ & 60.8$^{+5.3}_{-4.3}$ & 0.71$^{+0.06}_{-0.06}$ & 18.76 & BBH\\
			GWTC-3 & GW200128$\_$022011 & 0.56$^{+0.28}_{-0.28}$ & 42.2$^{+11.6}_{-8.1}$ & 32.6$^{+9.5}_{-9.2}$ & 71$^{+16}_{-11}$ & 0.74$^{+0.10}_{-0.10}$ & 19.11 & BBH\\
			GWTC-3 & GW200129$\_$065458 & 0.18$^{+0.05}_{-0.07}$ & 34.5$^{+9.9}_{-3.2}$ & 28.9$^{+3.4}_{-9.3}$ & 60.3$^{+4.0}_{-3.3}$ & 0.73$^{+0.06}_{-0.05}$ & 18.99 & BBH\\
			GWTC-3 & GW200208$\_$222617 & 0.66$^{+0.53}_{-0.29}$ & 51$^{+103}_{-30}$ & 12.3$^{+9.2}_{-5.5}$ & 61$^{+99}_{-26}$ & 0.83$^{+0.14}_{-0.27}$ & 19.14 & BBH\\
			GWTC-3 & GW200220$\_$061928 & 0.90$^{+0.55}_{-0.40}$ & 87$^{+40}_{-23}$ & 61$^{+26}_{-25}$ & 141$^{+51}_{-31}$ & 0.71$^{+0.15}_{-0.17}$ & 19.36 & BBH\\
			GWTC-3 & GW200224$\_$222234 & 0.32$^{+0.08}_{-0.11}$ & 40.0$^{+6.9}_{-4.5}$ & 32.5$^{+5.0}_{-7.2}$ & 68.6$^{+6.6}_{-4.7}$ & 0.73$^{+0.07}_{-0.07}$ & 19.21 & BBH\\
			GWTC-3 & GW200306$\_$093714 & 0.38$^{+0.24}_{-0.18}$ & 28.3$^{+17.1}_{-7.7}$ & 14.8$^{+6.5}_{-6.4}$ & 41.7$^{+12.3}_{-6.9}$ & 0.78$^{+0.11}_{-0.26}$ & 18.61 & BBH\\
			GWTC-3 & GW200308$\_$173609 & 1.04$^{+1.47}_{-0.57}$ & 60$^{+166}_{-29}$ & 24$^{+36}_{-13}$ & 88$^{+169}_{-47}$ & 0.91$^{+0.03}_{-0.08}$ & 20.20 & BBH\\
			GWTC-3 & GW200322$\_$091133 & 0.59$^{+1.43}_{-0.32}$ & 50$^{+132}_{-22}$ & 11.3$^{+24.3}_{-6.0}$ & 48$^{+132}_{-22}$ & 0.78$^{+0.16}_{-0.17}$ & 18.91 & BBH\\
			GWTC-4.0 & GW230814$\_$230901 & 0.06$^{+0.03}_{-0.03}$ & 33.6$^{+2.8}_{-2.2}$ & 28.3$^{+2.1}_{-3.0}$ & 58.9$^{+1.9}_{-1.9}$ & 0.71$^{+0.04}_{-0.05}$ & 19.36 & BBH\\
			GWTC-4.0 & GW231028$\_$153006 & 0.67$^{+0.18}_{-0.27}$ & 95.0$^{+33.0}_{-20.0}$ & 58.0$^{+21.0}_{-25.0}$ & 144.0$^{+27.0}_{-13.0}$ & 0.77$^{+0.08}_{-0.08}$ & 19.60 & BBH\\
			GWTC-4.0 & GW231226$\_$101520 & 0.23$^{+0.04}_{-0.06}$ & 40.1$^{+4.4}_{-2.9}$ & 35.0$^{+3.2}_{-4.9}$ & 71.4$^{+3.8}_{-2.8}$ & 0.73$^{+0.05}_{-0.04}$ & 19.11 & BBH\\
			\hline
			\hline
		\end{tabular}
	\end{table*}

	\section{Summary and conclusions}
	\label{sec:summary}
	
	We have investigated UHE particle production through the magnetized BSW mechanism in BH--NS binaries, BH--BH binaries in which at least one component has a charge-to-mass ratio of $\sim 10^{-4}$--$10^{-3}$, or post-merger BH remnants formed in binary NS mergers detected by LIGO-Virgo-KAGRA. By solving the geodesic equations for charged particles in magnetized Kerr spacetime, we demonstrate that binary coalescences represent viable astrophysical sources of ultra-high-energy cosmic rays.
	
	Our previous Letter \cite{2024PhRvL.132i1401P} demonstrated that magnetic fields exceeding approximately $10^{10}-10^{12}\,\mathrm{G}$ are sufficient to enable ultra-high-energy particle production in compact-binary environments. The present work extends that analysis by systematically investigating how the acceleration efficiency evolves as the magnetic field strength increases. In particular, we identify a transition toward a magnetic-dominated regime for fields approaching $10^{12}\,\mathrm{G}$ and quantify the resulting enhancement up to $B\sim10^{14}\,\mathrm{G}$. Therefore, the present results should be viewed as a refinement and extension of the original framework rather than a modification of its conclusions.
	
	The presence of strong magnetic fields in the immediate vicinity of the event horizon enhances the center-of-mass energy. It scales as $E_{\rm max} = m_0c^2(M/M_\odot)\mathcal{F}(\chi_f, \mathcal{B}, \ell_1, \ell_2)$, where the amplification function $\mathcal{F}$ depends on black hole spin $\chi_f$, normalized magnetic field strength $\mathcal{B}$, and particle angular momenta. For merger remnants with masses $M \sim 20$--$150\,M_\odot$ (characteristic of the BBH population), spins $\chi_f \sim 0.7$--$0.9$, and magnetic fields $B \sim 10^{12}$--$10^{14}$\,G, we find maximum achievable energies spanning $E_{\rm max} \sim 10^{17}$--$10^{20}$\,eV, placing these systems squarely in the UHECR regime.
	
	Our systematic parameter space exploration reveals three distinct acceleration regimes characterized by magnetic field strength, see Fig.~\ref{fig:magnetic_ratio}. In the gravity-dominated regime ($B < 10^{12}$ G), magnetic fields provide only modest enhancement, with amplification factors below $\sim 2$, and particle acceleration is driven primarily by spacetime curvature. As field strength increases into the transition regime ($10^{12}$ G $\lesssim B \lesssim 10^{13}$ G), magnetic and gravitational effects become comparable, with amplification factors rising steeply to $\sim 5$. This transition regime is particularly significant because it encompasses field strengths naturally expected in binary neutron star mergers, suggesting that magnetic enhancement of the BSW mechanism may be a generic feature rather than requiring fine-tuned conditions. In the magnetic-dominated regime ($B > 10^{13}$ G), corresponding to magnetar-strength fields, amplification factors approach $\sim 10$ and the magnetic field becomes the dominant driver of particle acceleration.
	
	A crucial finding is that magnetic fields dramatically broaden the viable parameter space compared to the original BSW scenario. While the vacuum BSW mechanism required extreme fine-tuning to near-extremal spins ($\chi_f \to 1$)~\cite{banados}, the presence of magnetic fields $B \gtrsim 10^{12}$ G enables UHE particle production across nearly the entire distribution of spins observed in gravitational wave catalogs ($\chi_f \gtrsim 0.7$). This eliminates the most severe astrophysical objection to the BSW mechanism and transforms it from a theoretical curiosity into a potentially realistic acceleration process operating in compact binary mergers.
	
	Applying our framework to the catalog of 34 high-spin gravitational wave events detected through GWTC-2, GWTC-3 and preliminary GWTC-4 observations, we find that systems with $\chi_f \gtrsim 0.85$ and $M \gtrsim 100\,M_\odot$ can reach $E_{\rm max} \sim 10^{20}$ eV at our fiducial magnetic field strength $B = 10^{14}$ G, while typical events with $\chi_f \sim 0.7$--$0.8$ and $M \sim 50$--$70\,M_\odot$ achieve $E_{\rm max} \sim 10^{18}$--$10^{19}$ eV, see Table~\ref{tab:gws}. These event-by-event predictions establish direct, testable correlations between gravitational wave observables and UHECR production efficiency.
	
	Our calculations adopt simplified assumptions including equatorial particle orbits and a uniform magnetic field configuration with $A_\phi = Bg_{\phi\phi}/2$. More realistic treatments incorporating full three-dimensional magnetospheric geometry from magnetohydrodynamic simulations may modify quantitative predictions while preserving the qualitative picture. The origin and magnitude of magnetic fields in binary black hole mergers remains uncertain, though our fiducial value $B = 10^{14}$ G represents a plausible upper estimate based on scaling arguments. Lower field strengths in the range $10^{12}$--$10^{13}$ G, more readily achieved through residual accretion or transient merger effects, still enable UHECR production for favorable parameter combinations. We have focused on collision energies near the horizon without detailed modeling of particle escape, propagation through the merger environment, or energy losses during transit to Earth. A complete assessment requires incorporating these effects along with realistic particle injection mechanisms and composition modeling.
	
	The prospects for testing this scenario through multi-messenger observations are promising. Third-generation gravitational wave detectors such as Cosmic Explorer~\cite{2019BAAS...51g..35R} and Einstein Telescope~\cite{2020JCAP...03..050M} will detect binary mergers with unprecedented precision, resolving remnant properties to percent-level accuracy and enabling detailed predictions for individual events. Simultaneously, next-generation cosmic ray observatories will improve mass composition measurements and directional resolution, potentially enabling coincident detection of UHECRs with gravitational wave sources~\cite{ANASTASI2022167497, olinto2021poemma, grand2018, deligny2022cosmic}. The detection or non-detection of UHECRs correlated with specific merger events would provide powerful constraints on magnetic field strengths and particle acceleration efficiency in these extreme environments.

	In conclusion, the present work should be viewed as a natural extension of our previous study \cite{2024PhRvL.132i1401P}. In that Letter, we demonstrated that astrophysically realistic magnetic fields larger than $\sim 10^{10}-10^{12}\,\mathrm{G}$ in the vicinity of black hole event horizons are already sufficient to trigger ultra-high-energy particle production. Here, we go beyond that initial result by identifying distinct acceleration regimes and quantifying the transition from gravity-dominated to magnetic-dominated dynamics, which occurs for magnetic fields $B \gtrsim 10^{12}\,\mathrm{G}$. These results do not alter the original conclusion; rather, they provide a deeper physical understanding of the mechanism and clarify how its efficiency evolves as the magnetic field strength increases. Taken together, the results of both studies strengthen the case that compact binary mergers embedded in strongly magnetized environments constitute theoretically robust and astrophysically plausible sources of ultra-high-energy cosmic rays. The inclusion of realistic magnetic fields transforms the BSW mechanism from requiring extreme fine-tuning to being a potentially generic feature of compact binary coalescences. Our quantitative predictions linking gravitational wave observables to particle acceleration efficiency provide a concrete framework for multi-messenger searches that could transform our understanding of the most energetic particles in the universe.
	
	\section{Acknowledgments}
	
	The authors acknowledge the support of the NAPI “Fenômenos Extremos do Universo” of Fundação de Apoio à Ciência, Tecnologia e Inovação do Paraná. C.H.C.-A. research is supported by Araucária Foundation (337/2025) and COFPI/PRPI/UFPR (20/2025). R.C.A. research is supported by CNPq (308859/2025-1) and (4000045/2023-0), Araucária Foundation (698/2022) and (721/2022) and FAPESP (2021/01089-1). The authors acknowledge the AWS Cloud Credit/CNPq and the National Laboratory for Scientific Computing (LNCC/MCTI, Brazil) for providing HPC resources of the SDumont supercomputer, which have contributed to the research results reported in this paper. URL: https://sdumont.lncc.br.  
	J.G.C. is
	grateful for the support of FAPES (1020/2022, 1081/2022, 976/2022,
	332/2023, 1514/2025), CNPq (311758/2021-5, 306018/2025-0), and FAPESP (grant
	No. 2021/01089-1).

	\bibliography{manuscript}

@PREAMBLE{
 "\providecommand{\noopsort}[1]{}" 
 # "\providecommand{\singleletter}[1]{#1}%" 
}

@ARTICLE{2025MNRAS.542.3067A,
       author = {{Aguilera-Miret}, Ricard and {Christian}, Jan-Erik and {Rosswog}, Stephan and {Palenzuela}, Carlos},
        title = "{Robustness of Magnetic Field Amplification in Neutron Star Mergers}",
      journal = {\mnras},
         year = 2025,
        month = aug,
       volume = {542},
       number = {4},
        pages = {3067-3077},
          doi = {10.1093/mnras/staf1291},
archivePrefix = {arXiv},
       eprint = {2504.10604},
 primaryClass = {astro-ph.HE},
       adsurl = {https://ui.adsabs.harvard.edu/abs/2025MNRAS.542.3067A}
}

@ARTICLE{2016ApJ...824L...6R,
       author = {{Ruiz}, Milton and {Lang}, Ryan N. and {Paschalidis}, Vasileios and {Shapiro}, Stuart L.},
        title = "{Binary Neutron Star Mergers: A Jet Engine for Short Gamma-Ray Bursts}",
      journal = {\apjl},
         year = 2016,
        month = jun,
       volume = {824},
       number = {1},
          eid = {L6},
        pages = {L6},
          doi = {10.3847/2041-8205/824/1/L6},
archivePrefix = {arXiv},
       eprint = {1604.02455},
 primaryClass = {astro-ph.HE},
       adsurl = {https://ui.adsabs.harvard.edu/abs/2016ApJ...824L...6R}
}

@ARTICLE{2015PhRvD..92f4034K,
       author = {{Kiuchi}, Kenta and {Sekiguchi}, Yuichiro and {Kyutoku}, Koutarou and {Shibata}, Masaru and {Taniguchi}, Keisuke and {Wada}, Tomohide},
        title = "{High resolution magnetohydrodynamic simulation of black hole-neutron star merger: Mass ejection and short gamma ray bursts}",
      journal = {\prd},
         year = 2015,
        month = sep,
       volume = {92},
       number = {6},
          eid = {064034},
        pages = {064034},
          doi = {10.1103/PhysRevD.92.064034},
archivePrefix = {arXiv},
       eprint = {1506.06811},
 primaryClass = {astro-ph.HE},
       adsurl = {https://ui.adsabs.harvard.edu/abs/2015PhRvD..92f4034K}
}

@ARTICLE{2020ARNPS..70...95R,
       author = {{Radice}, David and {Bernuzzi}, Sebastiano and {Perego}, Albino},
        title = "{The Dynamics of Binary Neutron Star Mergers and GW170817}",
      journal = {Annual Review of Nuclear and Particle Science},
         year = 2020,
        month = oct,
       volume = {70},
        pages = {95-119},
          doi = {10.1146/annurev-nucl-013120-114541},
archivePrefix = {arXiv},
       eprint = {2002.03863},
 primaryClass = {astro-ph.HE},
       adsurl = {https://ui.adsabs.harvard.edu/abs/2020ARNPS..70...95R}
}

@ARTICLE{2015ApJ...809...39G,
       author = {{Giacomazzo}, Bruno and {Zrake}, Jonathan and {Duffell}, Paul C. and {MacFadyen}, Andrew I. and {Perna}, Rosalba},
        title = "{Producing Magnetar Magnetic Fields in the Merger of Binary Neutron Stars}",
      journal = {\apj},
         year = 2015,
        month = aug,
       volume = {809},
       number = {1},
          eid = {39},
        pages = {39},
          doi = {10.1088/0004-637X/809/1/39},
archivePrefix = {arXiv},
       eprint = {1410.0013},
 primaryClass = {astro-ph.HE},
       adsurl = {https://ui.adsabs.harvard.edu/abs/2015ApJ...809...39G}
}

@ARTICLE{2010A&A...515A..30O,
       author = {{Obergaulinger}, M. and {Aloy}, M.~A. and {M{\"u}ller}, E.},
        title = "{Local simulations of the magnetized Kelvin-Helmholtz instability in neutron-star mergers}",
      journal = {\aap},
         year = 2010,
        month = jun,
       volume = {515},
          eid = {A30},
        pages = {A30},
          doi = {10.1051/0004-6361/200913386},
archivePrefix = {arXiv},
       eprint = {1003.6031},
 primaryClass = {astro-ph.SR},
       adsurl = {https://ui.adsabs.harvard.edu/abs/2010A&A...515A..30O}
}

@article{PhysRevD.106.023013,
  title = {Turbulent magnetic field amplification in binary neutron star mergers},
  author = {Palenzuela, Carlos and Aguilera-Miret, Ricard and Carrasco, Federico and Ciolfi, Riccardo and Kalinani, Jay Vijay and Kastaun, Wolfgang and Mi\~nano, Borja and Vigan\`o, Daniele},
  journal = {Phys. Rev. D},
  volume = {106},
  issue = {2},
  pages = {023013},
  numpages = {21},
  year = {2022},
  month = {Jul},
  publisher = {American Physical Society},
  doi = {10.1103/PhysRevD.106.023013},
  url = {https://link.aps.org/doi/10.1103/PhysRevD.106.023013}
}

@article{PierreAuger2020qqz,
    author = "Aab, Alexander and others",
    collaboration = "Pierre Auger",
    title = "{Measurement of the cosmic-ray energy spectrum above $2.5{\times} 10^{18}$  eV using the Pierre Auger Observatory}",
    eprint = "2008.06486",
    archivePrefix = "arXiv",
    primaryClass = "astro-ph.HE",
    reportNumber = "FERMILAB-PUB-20-432-E-TD",
    doi = "10.1103/PhysRevD.102.062005",
    journal = "Phys. Rev. D",
    volume = "102",
    number = "6",
    pages = "062005",
    year = "2020"
}

@ARTICLE{Bell1978,
       author = {{Bell}, A.~R.},
        title = "{The acceleration of cosmic rays in shock fronts - I.}",
      journal = {\mnras},
         year = 1978,
        month = jan,
       volume = {182},
        pages = {147-156},
          doi = {10.1093/mnras/182.2.147},
       adsurl = {https://ui.adsabs.harvard.edu/abs/1978MNRAS.182..147B}
}

@ARTICLE{2025PhRvL.134h1003F,
       author = {{Farrar}, Glennys R.},
        title = "{Binary Neutron Star Mergers as the Source of the Highest Energy Cosmic Rays}",
      journal = {\prl},
         year = 2025,
        month = feb,
       volume = {134},
       number = {8},
          eid = {081003},
        pages = {081003},
          doi = {10.1103/PhysRevLett.134.081003},
archivePrefix = {arXiv},
       eprint = {2405.12004},
 primaryClass = {astro-ph.HE},
       adsurl = {https://ui.adsabs.harvard.edu/abs/2025PhRvL.134h1003F}
}

@ARTICLE{2025ApJ...994L...7F,
       author = {{Farrar}, Glennys R.},
        title = "{Ultra-High-energy Cosmic Ray Production in Binary Neutron Star Mergers}",
      journal = {\apjl},
         year = 2025,
        month = nov,
       volume = {994},
       number = {1},
          eid = {L7},
        pages = {L7},
          doi = {10.3847/2041-8213/ae14d5},
archivePrefix = {arXiv},
       eprint = {2506.22625},
 primaryClass = {astro-ph.HE},
       adsurl = {https://ui.adsabs.harvard.edu/abs/2025ApJ...994L...7F}
}

@ARTICLE{ANCHORDOQUI20191,
       author = {{Anchordoqui}, Luis A.},
        title = "{Ultra-high-energy cosmic rays}",
      journal = {Physics Reports},
         year = 2019,
        month = apr,
       volume = {801},
        pages = {1-93},
          doi = {10.1016/j.physrep.2019.01.002},
archivePrefix = {arXiv},
       eprint = {1807.09645},
 primaryClass = {astro-ph.HE},
       adsurl = {https://ui.adsabs.harvard.edu/abs/2019PhR...801....1A}
}

@ARTICLE{ANASTASI2022167497,
       author = {{Anastasi}, Gioacchino Alex and {Pierre Auger Collaboration}},
        title = "{AugerPrime: The Pierre Auger Observatory upgrade}",
      journal = {Nuclear Instruments and Methods in Physics Research A},
         year = 2022,
        month = dec,
       volume = {1044},
          eid = {167497},
        pages = {167497},
          doi = {10.1016/j.nima.2022.167497},
       adsurl = {https://ui.adsabs.harvard.edu/abs/2022NIMPA104467497A}
}

@ARTICLE{1966JETPL...4...78Z,
       author = {{Zatsepin}, G.~T. and {Kuz'min}, V.~A.},
        title = "{Upper Limit of the Spectrum of Cosmic Rays}",
      journal = {Soviet Journal of Experimental and Theoretical Physics Letters},
         year = 1966,
        month = aug,
       volume = {4},
        pages = {78},
       adsurl = {https://ui.adsabs.harvard.edu/abs/1966JETPL...4...78Z}
}

@ARTICLE{PhysRevLett.16.748,
       author = {{Greisen}, Kenneth},
        title = "{End to the Cosmic-Ray Spectrum?}",
      journal = {\prl},
         year = 1966,
        month = apr,
       volume = {16},
       number = {17},
        pages = {748-750},
          doi = {10.1103/PhysRevLett.16.748},
       adsurl = {https://ui.adsabs.harvard.edu/abs/1966PhRvL..16..748G}
}

@ARTICLE{federico,
       author = {{Fraschetti}, Federico},
        title = "{On the acceleration of ultra-high-energy cosmic rays}",
      journal = {Philosophical Transactions of the Royal Society of London Series A},
         year = 2008,
        month = dec,
       volume = {366},
       number = {1884},
        pages = {4417-4428},
          doi = {10.1098/rsta.2008.0204},
archivePrefix = {arXiv},
       eprint = {0809.3057},
 primaryClass = {astro-ph},
       adsurl = {https://ui.adsabs.harvard.edu/abs/2008RSPTA.366.4417F}
}

@INPROCEEDINGS{2019BAAS...51g..35R,
       author = {{Reitze}, David and {Adhikari}, Rana X. and {Ballmer}, Stefan and {Barish}, Barry and {Barsotti}, Lisa and {Billingsley}, GariLynn and {Brown}, Duncan A. and {Chen}, Yanbei and {Coyne}, Dennis and {Eisenstein}, Robert and {Evans}, Matthew and {Fritschel}, Peter and {Hall}, Evan D. and {Lazzarini}, Albert and {Lovelace}, Geoffrey and {Read}, Jocelyn and {Sathyaprakash}, B.~S. and {Shoemaker}, David and {Smith}, Joshua and {Torrie}, Calum and {Vitale}, Salvatore and {Weiss}, Rainer and {Wipf}, Christopher and {Zucker}, Michael},
        title = "{Cosmic Explorer: The U.S. Contribution to Gravitational-Wave Astronomy beyond LIGO}",
    booktitle = {Bulletin of the American Astronomical Society},
         year = 2019,
       volume = {51},
        month = sep,
          eid = {35},
        pages = {35},
archivePrefix = {arXiv},
       eprint = {1907.04833},
 primaryClass = {astro-ph.IM},
       adsurl = {https://ui.adsabs.harvard.edu/abs/2019BAAS...51g..35R}
}

@ARTICLE{2020JCAP...03..050M,
       author = {{Maggiore}, Michele and {Van Den Broeck}, Chris and {Bartolo}, Nicola and {Belgacem}, Enis and {Bertacca}, Daniele and {Bizouard}, Marie Anne and {Branchesi}, Marica and {Clesse}, Sebastien and {Foffa}, Stefano and {Garc{\'\i}a-Bellido}, Juan and {Grimm}, Stefan and {Harms}, Jan and {Hinderer}, Tanja and {Matarrese}, Sabino and {Palomba}, Cristiano and {Peloso}, Marco and {Ricciardone}, Angelo and {Sakellariadou}, Mairi},
        title = "{Science case for the Einstein telescope}",
      journal = {\jcap},
         year = 2020,
        month = mar,
       volume = {2020},
       number = {3},
          eid = {050},
        pages = {050},
          doi = {10.1088/1475-7516/2020/03/050},
archivePrefix = {arXiv},
       eprint = {1912.02622},
 primaryClass = {astro-ph.CO},
       adsurl = {https://ui.adsabs.harvard.edu/abs/2020JCAP...03..050M}
}

@ARTICLE{kotera2011,
       author = {{Kotera}, Kumiko and {Olinto}, Angela V.},
        title = "{The Astrophysics of Ultrahigh-Energy Cosmic Rays}",
      journal = {\araa},
         year = 2011,
        month = sep,
       volume = {49},
       number = {1},
        pages = {119-153},
          doi = {10.1146/annurev-astro-081710-102620},
archivePrefix = {arXiv},
       eprint = {1101.4256},
 primaryClass = {astro-ph.HE},
       adsurl = {https://ui.adsabs.harvard.edu/abs/2011ARA&A..49..119K}
}

@ARTICLE{ta,
       author = {{Abu-Zayyad}, T. and others},
        title = "{The Cosmic-Ray Energy Spectrum Observed with the Surface Detector of the Telescope Array Experiment}",
      journal = {\apjl},
         year = 2013,
        month = may,
       volume = {768},
       number = {1},
          eid = {L1},
        pages = {L1},
          doi = {10.1088/2041-8205/768/1/L1},
archivePrefix = {arXiv},
       eprint = {1205.5067},
 primaryClass = {astro-ph.HE},
       adsurl = {https://ui.adsabs.harvard.edu/abs/2013ApJ...768L...1A}
}

@INPROCEEDINGS{coimbra2018,
       author = {{Coimbra Araujo}, C.~H. and {Anjos}, R.~C.},
        title = "{Rotating black holes with magnetic fields as accelerators of charged particles}",
    booktitle = {International Conference on Black Holes as Cosmic Batteries: UHECRs and Multimessenger Astronomy. 12-15 September 2018. Foz do Igua{\c{c}}u},
         year = 2018,
        month = sep,
          eid = {5},
        pages = {5},
          doi = {10.22323/1.329.0005},
       adsurl = {https://ui.adsabs.harvard.edu/abs/2018bhcb.confE...5C}
}

@ARTICLE{coimbra2020,
       author = {{Coimbra-Ara{\'u}jo}, C.~H. and {Anjos}, R.~C.},
        title = "{Acceleration of charged particles from near-extremal rotating black holes embedded in magnetic fields}",
      journal = {Classical and Quantum Gravity},
         year = 2021,
        month = jan,
       volume = {38},
       number = {1},
          eid = {015007},
        pages = {015007},
          doi = {10.1088/1361-6382/abc189},
archivePrefix = {arXiv},
       eprint = {2005.13599},
 primaryClass = {gr-qc},
       adsurl = {https://ui.adsabs.harvard.edu/abs/2021CQGra..38a5007C}
}

@ARTICLE{coimbra2022,
       author = {{Coimbra-Ara{\'u}jo}, Carlos H. and {dos Anjos}, Rita C.},
        title = "{Ultra-High-Energy Particles at the Border of Kerr Black Holes Triggered by Magnetocentrifugal Winds}",
      journal = {Galaxies},
         year = 2022,
        month = jul,
       volume = {10},
       number = {4},
        pages = {84},
          doi = {10.3390/galaxies10040084},
       adsurl = {https://ui.adsabs.harvard.edu/abs/2022Galax..10...84C}
}

@ARTICLE{2017ApJ...848L..12A,
       author = {{Abbott}, B.~P. and others},
        title = "{Multi-messenger Observations of a Binary Neutron Star Merger}",
      journal = {\apjl},
         year = 2017,
        month = oct,
       volume = {848},
       number = {2},
          eid = {L12},
        pages = {L12},
          doi = {10.3847/2041-8213/aa91c9},
       adsurl = {https://ui.adsabs.harvard.edu/abs/2017ApJ...848L..12A}
}

@ARTICLE{2017ApJ...850L..19M,
       author = {{Margalit}, Ben and {Metzger}, Brian D.},
        title = "{Constraining the Maximum Mass of Neutron Stars from Multi-messenger Observations of GW170817}",
      journal = {\apjl},
         year = 2017,
        month = dec,
       volume = {850},
       number = {2},
          eid = {L19},
        pages = {L19},
          doi = {10.3847/2041-8213/aa991c},
       adsurl = {https://ui.adsabs.harvard.edu/abs/2017ApJ...850L..19M}
}

@ARTICLE{2019ARNPS..69...41S,
       author = {{Shibata}, Masaru and {Hotokezaka}, Kenta},
        title = "{Merger and Mass Ejection of Neutron-Star Binaries}",
      journal = {Annual Review of Nuclear and Particle Science},
         year = 2019,
        month = oct,
       volume = {69},
        pages = {41-64},
          doi = {10.1146/annurev-nucl-101918-023625},
       adsurl = {https://ui.adsabs.harvard.edu/abs/2019ARNPS..69...41S}
}

@article{PhysRevD.97.124039,
  title = {Global simulations of strongly magnetized remnant massive neutron stars formed in binary neutron star mergers},
  author = {Kiuchi, Kenta and Kyutoku, Koutarou and Sekiguchi, Yuichiro and Shibata, Masaru},
  journal = {Phys. Rev. D},
  volume = {97},
  issue = {12},
  pages = {124039},
  numpages = {16},
  year = {2018},
  month = {Jun},
  publisher = {American Physical Society},
  doi = {10.1103/PhysRevD.97.124039},
  url = {https://link.aps.org/doi/10.1103/PhysRevD.97.124039}
}

@ARTICLE{2012PhRvD..85b4020F,
       author = {{Frolov}, Valeri P.},
        title = "{Weakly magnetized black holes as particle accelerators}",
      journal = {\prd},
         year = 2012,
        month = jan,
       volume = {85},
       number = {2},
          eid = {024020},
        pages = {024020},
          doi = {10.1103/PhysRevD.85.024020},
       adsurl = {https://ui.adsabs.harvard.edu/abs/2012PhRvD..85b4020F}
}

@ARTICLE{2010PhRvD..82h3004Z,
       author = {{Zaslavskii}, O.~B.},
        title = "{Acceleration of particles by black holes: General considerations}",
      journal = {\prd},
         year = 2010,
        month = oct,
       volume = {82},
       number = {8},
          eid = {083004},
        pages = {083004},
          doi = {10.1103/PhysRevD.82.083004},
       adsurl = {https://ui.adsabs.harvard.edu/abs/2010PhRvD..82h3004Z}
}

@ARTICLE{2014A&A...562A.137F,
       author = {{Falcke}, Heino and {Rezzolla}, Luciano},
        title = "{Fast radio bursts: the last crack of doom}",
      journal = {\aap},
         year = 2014,
        month = feb,
       volume = {562},
          eid = {A137},
        pages = {A137},
          doi = {10.1051/0004-6361/201321996},
       adsurl = {https://ui.adsabs.harvard.edu/abs/2014A&A...562A.137F}
}

@ARTICLE{2015MPLA...3050076Z,
       author = {{Zaslavskii}, O.~B.},
        title = "{Unbounded energies of debris from head-on particle collisions near black holes}",
      journal = {Modern Physics Letters A},
         year = 2015,
        month = may,
       volume = {30},
       number = {16},
          eid = {1550076},
        pages = {1550076},
          doi = {10.1142/S0217732315500765},
archivePrefix = {arXiv},
       eprint = {1411.0267},
 primaryClass = {gr-qc},
       adsurl = {https://ui.adsabs.harvard.edu/abs/2015MPLA...3050076Z}
}

@ARTICLE{2014MPLA...2950112Z,
       author = {{Zaslavskii}, O.~B.},
        title = "{Ultrahigh energy particle collisions near the black hole horizon in the strong magnetic field}",
      journal = {Modern Physics Letters A},
         year = 2014,
        month = jun,
       volume = {29},
       number = {21},
          eid = {1450112},
        pages = {1450112},
          doi = {10.1142/S0217732314501120},
archivePrefix = {arXiv},
       eprint = {1403.6286},
 primaryClass = {gr-qc},
       adsurl = {https://ui.adsabs.harvard.edu/abs/2014MPLA...2950112Z}
}

@ARTICLE{2015PhRvL.114y1103B,
       author = {{Berti}, Emanuele and {Brito}, Richard and {Cardoso}, Vitor},
        title = "{Ultrahigh-Energy Debris from the Collisional Penrose Process}",
      journal = {\prl},
         year = 2015,
        month = jun,
       volume = {114},
       number = {25},
          eid = {251103},
        pages = {251103},
          doi = {10.1103/PhysRevLett.114.251103},
archivePrefix = {arXiv},
       eprint = {1410.8534},
 primaryClass = {gr-qc},
       adsurl = {https://ui.adsabs.harvard.edu/abs/2015PhRvL.114y1103B}
}

@article{PhysRevD.86.024027,
  title = {Upper limits of particle emission from high-energy collision and reaction near a maximally rotating Kerr black hole},
  author = {Harada, Tomohiro and Nemoto, Hiroya and Miyamoto, Umpei},
  journal = {Phys. Rev. D},
  volume = {86},
  issue = {2},
  pages = {024027},
  numpages = {10},
  year = {2012},
  month = {Jul},
  publisher = {American Physical Society},
  doi = {10.1103/PhysRevD.86.024027},
  url = {https://link.aps.org/doi/10.1103/PhysRevD.86.024027}
}

@ARTICLE{2012PhRvD..86h4030Z,
       author = {{Zaslavskii}, O.~B.},
        title = "{Energetics of particle collisions near dirty rotating extremal black holes: Banados-Silk-West effect versus Penrose process}",
      journal = {\prd},
         year = 2012,
        month = oct,
       volume = {86},
       number = {8},
          eid = {084030},
        pages = {084030},
          doi = {10.1103/PhysRevD.86.084030},
archivePrefix = {arXiv},
       eprint = {1205.4410},
 primaryClass = {gr-qc},
       adsurl = {https://ui.adsabs.harvard.edu/abs/2012PhRvD..86h4030Z}
}

@ARTICLE{2012PhRvL.109l1101B,
       author = {{Bejger}, Micha{\l} and {Piran}, Tsvi and {Abramowicz}, Marek and {H{\r{a}}kanson}, Frida},
        title = "{Collisional Penrose Process near the Horizon of Extreme Kerr Black Holes}",
      journal = {\prl},
         year = 2012,
        month = sep,
       volume = {109},
       number = {12},
          eid = {121101},
        pages = {121101},
          doi = {10.1103/PhysRevLett.109.121101},
archivePrefix = {arXiv},
       eprint = {1205.4350},
 primaryClass = {astro-ph.HE},
       adsurl = {https://ui.adsabs.harvard.edu/abs/2012PhRvL.109l1101B}
}

@ARTICLE{banados,
       author = {{Ba{\~n}ados}, M{\'a}ximo and {Silk}, Joseph and {West}, Stephen M.},
        title = "{Kerr Black Holes as Particle Accelerators to Arbitrarily High Energy}",
      journal = {\prl},
         year = 2009,
        month = sep,
       volume = {103},
       number = {11},
          eid = {111102},
        pages = {111102},
          doi = {10.1103/PhysRevLett.103.111102},
archivePrefix = {arXiv},
       eprint = {0909.0169},
 primaryClass = {hep-ph},
       adsurl = {https://ui.adsabs.harvard.edu/abs/2009PhRvL.103k1102B}
}

@ARTICLE{jacobson,
       author = {{Jacobson}, Ted and {Sotiriou}, Thomas P.},
        title = "{Spinning Black Holes as Particle Accelerators}",
      journal = {\prl},
         year = 2010,
        month = jan,
       volume = {104},
       number = {2},
          eid = {021101},
        pages = {021101},
          doi = {10.1103/PhysRevLett.104.021101},
archivePrefix = {arXiv},
       eprint = {0911.3363},
 primaryClass = {gr-qc},
       adsurl = {https://ui.adsabs.harvard.edu/abs/2010PhRvL.104b1101J}
}

@ARTICLE{wei,
       author = {{Wei}, Shao-Wen and {Liu}, Yu-Xiao and {Guo}, Heng and {Fu}, Chun-E.},
        title = "{Charged spinning black holes as particle accelerators}",
      journal = {\prd},
         year = 2010,
        month = nov,
       volume = {82},
       number = {10},
          eid = {103005},
        pages = {103005},
          doi = {10.1103/PhysRevD.82.103005},
archivePrefix = {arXiv},
       eprint = {1006.1056},
 primaryClass = {hep-th},
       adsurl = {https://ui.adsabs.harvard.edu/abs/2010PhRvD..82j3005W}
}

@ARTICLE{frolov,
       author = {{Frolov}, Valeri P.},
        title = "{Weakly magnetized black holes as particle accelerators}",
      journal = {\prd},
         year = 2012,
        month = jan,
       volume = {85},
       number = {2},
          eid = {024020},
        pages = {024020},
          doi = {10.1103/PhysRevD.85.024020},
archivePrefix = {arXiv},
       eprint = {1110.6274},
 primaryClass = {gr-qc},
       adsurl = {https://ui.adsabs.harvard.edu/abs/2012PhRvD..85b4020F}
}

@ARTICLE{igata,
       author = {{Igata}, Takahisa and {Harada}, Tomohiro and {Kimura}, Masashi},
        title = "{Effect of a weak electromagnetic field on particle acceleration by a rotating black hole}",
      journal = {\prd},
         year = 2012,
        month = may,
       volume = {85},
       number = {10},
          eid = {104028},
        pages = {104028},
          doi = {10.1103/PhysRevD.85.104028},
archivePrefix = {arXiv},
       eprint = {1202.4859},
 primaryClass = {gr-qc},
       adsurl = {https://ui.adsabs.harvard.edu/abs/2012PhRvD..85j4028I}
}

@ARTICLE{press,
       author = {{Bardeen}, James M. and {Press}, William H. and {Teukolsky}, Saul A.},
        title = "{Rotating Black Holes: Locally Nonrotating Frames, Energy Extraction, and Scalar Synchrotron Radiation}",
      journal = {\apj},
         year = 1972,
        month = dec,
       volume = {178},
        pages = {347-370},
          doi = {10.1086/151796},
       adsurl = {https://ui.adsabs.harvard.edu/abs/1972ApJ...178..347B}
}

@ARTICLE{2012PhRvD..85l4062Z,
       author = {{Zilh{\~a}o}, Miguel and {Cardoso}, Vitor and {Herdeiro}, Carlos and {Lehner}, Luis and {Sperhake}, Ulrich},
        title = "{Collisions of charged black holes}",
      journal = {\prd},
         year = 2012,
        month = jun,
       volume = {85},
       number = {12},
          eid = {124062},
        pages = {124062},
          doi = {10.1103/PhysRevD.85.124062},
archivePrefix = {arXiv},
       eprint = {1205.1063},
 primaryClass = {gr-qc},
       adsurl = {https://ui.adsabs.harvard.edu/abs/2012PhRvD..85l4062Z}
}

@ARTICLE{schn,
       author = {{Schnittman}, Jeremy D.},
        title = "{Revised Upper Limit to Energy Extraction from a Kerr Black Hole}",
      journal = {\prl},
         year = 2014,
        month = dec,
       volume = {113},
       number = {26},
          eid = {261102},
        pages = {261102},
          doi = {10.1103/PhysRevLett.113.261102},
archivePrefix = {arXiv},
       eprint = {1410.6446},
 primaryClass = {astro-ph.HE},
       adsurl = {https://ui.adsabs.harvard.edu/abs/2014PhRvL.113z1102S}
}

@ARTICLE{PhysRevLett.134.021001,
       author = {{Halim}, A. and others},
       collaboration = {Pierre Auger Collaboration},
        title = "{Inference of the Mass Composition of Cosmic Rays with Energies from 1018.5 to 1020  eV Using the Pierre Auger Observatory and Deep Learning}",
      journal = {\prl},
         year = 2025,
        month = jan,
       volume = {134},
       number = {2},
          eid = {021001},
        pages = {021001},
          doi = {10.1103/PhysRevLett.134.021001},
archivePrefix = {arXiv},
       eprint = {2406.06315},
 primaryClass = {astro-ph.HE},
       adsurl = {https://ui.adsabs.harvard.edu/abs/2025PhRvL.134b1001A}
}

@ARTICLE{2025arXiv250710292M,
       author = {{Mayotte}, Eric},
        title = "{Measurement and Interpretation of UHECR Mass Composition at the Pierre Auger Observatory}",
      journal = {arXiv e-prints},
         year = 2025,
        month = jul,
          eid = {arXiv:2507.10292},
        pages = {arXiv:2507.10292},
          doi = {10.48550/arXiv.2507.10292},
archivePrefix = {arXiv},
       eprint = {2507.10292},
 primaryClass = {astro-ph.HE},
       adsurl = {https://ui.adsabs.harvard.edu/abs/2025arXiv250710292M}
}

@article{gwtc2,
  title = {{GWTC-2: Compact Binary Coalescences Observed by LIGO and Virgo during the First Half of the Third Observing Run}},
  author = {Abbott, R. and others},
  collaboration = {LIGO Scientific Collaboration and Virgo Collaboration},
  journal = {Phys. Rev. X},
  volume = {11},
  pages = {021053},
  year = {2021},
  doi = {10.1103/PhysRevX.11.021053},
  eprint = {2010.14527},
  archivePrefix = {arXiv},
  primaryClass = {gr-qc}
}

@article{gwtc3,
  title = {{GWTC-3: Compact Binary Coalescences Observed by LIGO and Virgo during the Second Part of the Third Observing Run}},
  author = {Abbott, R. and others},
  collaboration = {LIGO Scientific Collaboration and Virgo Collaboration},
  journal = {Phys. Rev. X},
  volume = {13},
  pages = {041039},
  year = {2023},
  doi = {10.1103/PhysRevX.13.041039},
  eprint = {2111.03606},
  archivePrefix = {arXiv},
  primaryClass = {gr-qc}
}

@article{abbott2023population,
  title = {{Population of Merging Compact Binaries Inferred Using Gravitational Waves through GWTC-3}},
  author = {Abbott, R. and others},
  collaboration = {LIGO Scientific Collaboration, Virgo Collaboration, and KAGRA Collaboration},
  journal = {Phys. Rev. X},
  volume = {13},
  pages = {011048},
  year = {2023},
  doi = {10.1103/PhysRevX.13.011048},
  eprint = {2111.03634},
  archivePrefix = {arXiv},
  primaryClass = {astro-ph.HE}
}

@ARTICLE{2020GReGr..52...59C,
       author = {{Ciolfi}, Riccardo},
        title = "{The key role of magnetic fields in binary neutron star mergers}",
      journal = {General Relativity and Gravitation},
         year = 2020,
        month = jun,
       volume = {52},
       number = {6},
          eid = {59},
        pages = {59},
          doi = {10.1007/s10714-020-02714-x},
archivePrefix = {arXiv},
       eprint = {2003.07572},
 primaryClass = {astro-ph.HE},
       adsurl = {https://ui.adsabs.harvard.edu/abs/2020GReGr..52...59C}
}

@ARTICLE{2024PhRvL.132i1401P,
       author = {{Pereira}, Jonas P. and {Coimbra-Ara{\'u}jo}, Carlos H. and {dos Anjos}, Rita C. and {Coelho}, Jaziel G.},
        title = "{Binary Coalescences as Sources of Ultrahigh-Energy Cosmic Rays}",
      journal = {\prl},
         year = 2024,
        month = feb,
       volume = {132},
       number = {9},
          eid = {091401},
        pages = {091401},
          doi = {10.1103/PhysRevLett.132.091401},
archivePrefix = {arXiv},
       eprint = {2307.06200},
 primaryClass = {astro-ph.HE},
       adsurl = {https://ui.adsabs.harvard.edu/abs/2024PhRvL.132i1401P}
}

@article{olinto2021poemma,
  title = {The POEMMA (Probe of Extreme Multi-Messenger Astrophysics) observatory},
  author = {Olinto, A. V. and Krizmanic, J. and Adams, J. H. and Aloisio, R. and Anchordoqui, L. A. and Anzalone, A. and Bagheri, M. and Barghini, D. and Battisti, M. and Bergman, D. R. and others},
  journal = {J. Cosmology Astropart. Phys.},
  volume = {2021},
  number = {06},
  pages = {007},
  year = {2021},
  doi = {10.1088/1475-7516/2021/06/007},
  eprint = {2012.07945},
  archivePrefix = {arXiv},
  primaryClass = {astro-ph.HE}
}

@article{grand2018,
  title = {The Giant Radio Array for Neutrino Detection (GRAND): Science and Design},
  author = {{GRAND Collaboration} and Alvarez-Mu{\~n}iz, J. and Alves Batista, R. and Balagopal V., A. and Bon{\v c}ius, J. and Bustamante, M. and Carvalho, W. and Charrier, D. and Cognard, I. and Decoene, V. and others},
  journal = {Science China Physics, Mechanics, and Astronomy},
  volume = {63},
  pages = {219501},
  year = {2020},
  doi = {10.1007/s11433-018-9385-7},
  eprint = {1810.09994},
  archivePrefix = {arXiv},
  primaryClass = {astro-ph.HE}
}

@article{deligny2022cosmic,
  title = {Cosmic-Ray Extremely Distributed Observatory},
  author = {de L{\'e}on, S. Cuoco and others},
  collaboration = {CREDO Collaboration},
  journal = {Symmetry},
  volume = {12},
  number = {11},
  pages = {1835},
  year = {2020},
  doi = {10.3390/sym12111835}
}

@ARTICLE{1999APh....10..185P,
       author = {{Protheroe}, R.~J. and {Stanev}, Todor},
        title = "{Cut-offs and pile-ups in shock acceleration spectra}",
      journal = {Astroparticle Physics},
         year = 1999,
        month = mar,
       volume = {10},
       number = {2-3},
        pages = {185-196},
          doi = {10.1016/S0927-6505(98)00055-3},
archivePrefix = {arXiv},
       eprint = {astro-ph/9808129},
 primaryClass = {astro-ph},
       adsurl = {https://ui.adsabs.harvard.edu/abs/1999APh....10..185P}
}

@ARTICLE{2025MNRAS.539.2435E,
       author = {{Ehlert}, Domenik and {Oikonomou}, Foteini and {Peretti}, Enrico},
        title = "{Ultra-high-energy cosmic rays from ultra-fast outflows of active galactic nuclei}",
      journal = {\mnras},
         year = 2025,
        month = may,
       volume = {539},
       number = {3},
        pages = {2435-2462},
          doi = {10.1093/mnras/staf457},
archivePrefix = {arXiv},
       eprint = {2411.05667},
 primaryClass = {astro-ph.HE},
       adsurl = {https://ui.adsabs.harvard.edu/abs/2025MNRAS.539.2435E}
}

@article{bell2013,
  title = {Cosmic ray acceleration},
  author = {Bell, A. R.},
  journal = {Astroparticle Physics},
  volume = {43},
  pages = {56--70},
  year = {2013},
  doi = {10.1016/j.astropartphys.2012.05.022},
  eprint = {1301.7041},
  archivePrefix = {arXiv},
  primaryClass = {astro-ph.HE}
}

@article{spitkovsky2008,
  title = {Particle Acceleration in Relativistic Collisionless Shocks: Fermi Process at Last?},
  author = {Spitkovsky, A.},
  journal = {Astrophys. J. Lett.},
  volume = {682},
  pages = {L5},
  year = {2008},
  doi = {10.1086/590248},
  eprint = {0802.3216},
  archivePrefix = {arXiv},
  primaryClass = {astro-ph}
}

@article{caprioli2014,
  title = {Simulations of Ion Acceleration at Non-relativistic Shocks. I. Acceleration Efficiency},
  author = {Caprioli, D. and Spitkovsky, A.},
  journal = {Astrophys. J.},
  volume = {783},
  pages = {91},
  year = {2014},
  doi = {10.1088/0004-637X/783/2/91},
  eprint = {1310.2943},
  archivePrefix = {arXiv},
  primaryClass = {astro-ph.HE}
}

@ARTICLE{2025JCoPh.52313690S,
       author = {{Schween}, Nils W. and {Schulze}, Florian and {Reville}, Brian},
        title = "{Sapphire++: A particle transport code combining a spherical harmonic expansion and the discontinuous Galerkin method}",
      journal = {Journal of Computational Physics},
         year = 2025,
        month = feb,
       volume = {523},
          eid = {113690},
        pages = {113690},
          doi = {10.1016/j.jcp.2024.113690},
archivePrefix = {arXiv},
       eprint = {2501.05110},
 primaryClass = {astro-ph.HE},
       adsurl = {https://ui.adsabs.harvard.edu/abs/2025JCoPh.52313690S}
}

@ARTICLE{2025MNRAS.544L.160S,
       author = {{Shirin T}, Asma and {Reville}, Brian and {Schween}, Nils W. and {Schulze}, Florian and {Kirk}, John G.},
        title = "{Spectral curvature and breaks from Fermi acceleration at oblique shocks}",
      journal = {\mnras},
         year = 2025,
        month = nov,
       volume = {544},
       number = {1},
        pages = {L160-L166},
          doi = {10.1093/mnrasl/slaf113},
archivePrefix = {arXiv},
       eprint = {2511.01635},
 primaryClass = {astro-ph.HE},
       adsurl = {https://ui.adsabs.harvard.edu/abs/2025MNRAS.544L.160S}
}

@article{Berti2009,
  title = {Quasinormal modes of black holes and black branes},
  author = {Berti, Emanuele and Cardoso, Vitor and Starinets, Andrei O.},
  journal = {Classical and Quantum Gravity},
  volume = {26},
  number = {16},
  pages = {163001},
  year = {2009},
  publisher = {IOP Publishing},
  doi = {10.1088/0264-9381/26/16/163001},
  eprint = {0905.2975},
  archivePrefix = {arXiv},
  primaryClass = {gr-qc}
}

@article{Leaver1985,
  title = {An analytic representation for the quasi-normal modes of Kerr black holes},
  author = {Leaver, E. W.},
  journal = {Proceedings of the Royal Society of London. A. Mathematical and Physical Sciences},
  volume = {402},
  number = {1823},
  pages = {285--313},
  year = {1985},
  publisher = {The Royal Society London},
  doi = {10.1098/rspa.1985.0119}
}

@ARTICLE{2021PhRvD.103f4007F,
       author = {{Foucart}, Francois and {Chernoglazov}, Alexander and {Boyle}, Michael and {Hinderer}, Tanja and {Miller}, Max and {Moxon}, Jordan and {Scheel}, Mark A. and {Deppe}, Nils and {Duez}, Matthew D. and {H{\'e}bert}, Francois and {Kidder}, Lawrence E. and {Throwe}, William and {Pfeiffer}, Harald P.},
        title = "{High-accuracy waveforms for black hole-neutron star systems with spinning black holes}",
      journal = {\prd},
         year = 2021,
        month = mar,
       volume = {103},
       number = {6},
          eid = {064007},
        pages = {064007},
          doi = {10.1103/PhysRevD.103.064007},
archivePrefix = {arXiv},
       eprint = {2010.14518},
 primaryClass = {gr-qc},
       adsurl = {https://ui.adsabs.harvard.edu/abs/2021PhRvD.103f4007F}
}

@ARTICLE{2026ApJ..1002...97P,
       author = {{Padilha}, Luana N. and {Dos Anjos}, R.~C.},
        title = "{Oblique Shocks at Supernova Remnants in Massive Star Clusters: A Model for the Cosmic-Ray Knee Observed by LHAASO}",
      journal = {\apj},
         year = 2026,
        month = may,
       volume = {1002},
       number = {1},
          eid = {97},
        pages = {97},
          doi = {10.3847/1538-4357/ae5b95},
archivePrefix = {arXiv},
       eprint = {2604.07977},
 primaryClass = {astro-ph.HE},
       adsurl = {https://ui.adsabs.harvard.edu/abs/2026ApJ..1002...97P}
}
	
\end{document}